\documentclass[paper,11pt]{JHEP}
\usepackage[centertags]{amsmath}
\usepackage{amsfonts} \usepackage{amssymb} \usepackage{amsthm}
\usepackage{graphicx}

\newcommand{\eq}[1]{Eq.~(\ref{#1})}
\newcommand{\comm}[2]{\left[#1,#2\right]}
\newcommand{\bigcomm}[2]{\Bigl[#1,#2\Bigr]}
\newcommand{\acomm}[2]{\left\{#1,#2\right\}}

\newcommand{\ket}[1]{|{#1}\rangle}

\newcommand{\ii}{\mathrm{i}}
\newcommand{\ee}{\mathrm{e}}
\newcommand{\one}{{\rm 1\kern -.9mm l}}

\newcommand{\lvev}{\Big\langle\hskip -5pt\Big\langle}
\newcommand{\rvev}{\Big\rangle\hskip -5pt\Big\rangle}

\newcommand{\Tr}{\mathrm{Tr}\,}
\newcommand{\tr}{\mathrm{tr}\,}

\newcommand{\im}{\mbox{Im}\,}
\newcommand{\cF}{{\mathcal{F}}}
\newcommand{\bcF}{\bar{\mathcal{F}}}
\newcommand{\cE}{{\mathcal{E}}}
\newcommand{\cM}{{\mathcal{M}}}
\newcommand{\cN}{{\mathcal{N}}}
\newcommand{\cP}{{\mathcal{P}}}
\newcommand{\cQ}{{\mathcal{Q}}}
\newcommand{\cR}{{\mathcal{R}}}
\newcommand{\cS}{{\mathcal{S}}}
\newcommand{\cT}{{\mathcal{T}}}
\newcommand{\cW}{{\mathcal{W}}}
\newcommand{\Th}{T_{\rm h}}
\newcommand{\Uh}{U_{\rm h}}
\newcommand{\qh}{q_{\rm h}}
\newcommand{\Pf}{\mathrm{Pf}\,}
\newcommand{\ve}[1]{{\vec e}_{#1}}
\newcommand{\soq}{\big[\mathrm{SO}(8)\big]^4}

\title{Exotic instanton counting and heterotic/type I$^\prime$ duality}
\author{\parbox{13.5cm}{Marco Bill\`o$^1$, Livia Ferro$^2$, Marialuisa Frau$^1$,
Laurent Gallot$^2$, Alberto Lerda$^3$ and Igor Pesando$^1$}
\\
~\\
~\\
$^1$Dipartimento di Fisica Teorica, Universit\`a di Torino\\
and I.N.F.N. - sezione di Torino \\
Via P. Giuria 1, I-10125 Torino, Italy\\
\vspace{0.3cm}
$^2$
LAPTH, Universit\'e de Savoie, CNRS\\
9, chemin de Bellevue, BP 110, 74941 Annecy le Vieux Cedex, France\\
\vspace{0.3cm}
$^3$Dipartimento di Scienze e Tecnologie Avanzate, Universit\`a del Piemonte
Orientale\\
and I.N.F.N. - Gruppo Collegato di Alessandria - sezione di Torino\\
Viale T. Michel  11, I-15121 Alessandria, Italy\\
\vspace{0.3cm}
\email{billo,frau,lerda,ipesando@to.infn.it;
livia.ferro,laurent.gallot@lapp.in2p3.fr }
}

\abstract{We compute the partition function for the exotic instanton
system corresponding to D-instantons on D7 branes in Type I$^\prime$ theory. We
exploit the BRST structure of the moduli action and its deformation by
RR background to fully localize the integration. The resulting prepotential
describes non-perturbative corrections to the quartic couplings of the
gauge field $F$ living on the D7's. The results
match perfectly those obtained in the dual heterotic theory from 
a protected 1-loop computation, thus providing a non-trivial test of the duality
itself.}
\preprint{DFTT/41/2009\\LAPTH 1329/09}

\begin{document}

\section{Introduction and motivations}
\label{sec:intro}
In recent years, the possibility of acquiring some control over space-time
non-perturbative effects has been a unifying theme behind many developments in
String Theory. 

Much progress in this direction has been realized by exploiting the web of
dualities relating the five 10-dimensional string theories and the
11-dimensional M-theory through operations that map classical or perturbative
statements in one model to non-perturbative statements in its dual. One of the
most notable examples of such relations is the heterotic/type~I duality which
has been tested by checking the stable spectra on both sides
\cite{Polchinski:1995df}, and by studying   the BPS-saturated quartic couplings
$F^4$ for the gauge field and their gravitational counterparts $F^2 \mathcal
R^2$ or $\mathcal R^4$
\cite{Tseytlin:1995fy}-\nocite{Tseytlin:1995bi,Bachas:1996bp,
Bachas:1997mc,Kiritsis:1997hf,Bachas:1997xn,Lerche:1998nx,Foerger:1998kw,
Gutperle:1999xu,Gava:1999ky} \cite{Kiritsis:2000zi}~\footnote{For earlier
calculations of higher order couplings in the heterotic theory see
Ref.s~\cite{Lerche:1987sg}-\nocite{Lerche:1987qk}\cite{Lerche:1988zy}.}. In this
context, these protected quartic interactions are completely captured by a
1-loop computation on the heterotic side, while on the type I side they receive
both perturbative and non-perturbative contributions. 

In this paper we will consider a set-up in which the heterotic theory is
compactified on a 2-torus $\mathcal T_2$ with Wilson lines breaking the gauge
group to $\big[\mathrm{SO}(8)\big]^4$. In the dual theory, called type
I$^\prime$, the gauge degrees of freedom are supported by stacks of D7-branes,
while the non-perturbative contributions arise by adding D(--1)-branes, also
called D-instantons. Using the recent advances in the instanton calculus in
string theory (for a review see \cite{Blumenhagen:2009qh}) together with
localization techniques \cite{Moore:1998et,Nekrasov:2002qd}, we will extract
from the integration over the D-instanton moduli the quartic type I$^\prime$
interactions for the gauge fields and their gravitational corrections, and check
explicitly (up to instanton number $k=5$) the agreement with the heterotic
expressions. Although the structure of the type I$^\prime$ contributions has
already been investigated in the literature
\cite{Lerche:1998nx}-\nocite{Foerger:1998kw,Gutperle:1999xu,Gava:1999ky}\cite{
Kiritsis:2000zi}, and checks of the heterotic results have been performed
against the F-theory background that should represent the non-perturbative
completion of the type I$^\prime$ model \cite{Sen:1996vd}, we think that our
calculations provide the first case in which the agreement is verified by a
direct explicit evaluation of non-perturbative corrections in the
``microscopic'' theory.

The heterotic/type I$^\prime$ duality is not the only motivation for the
computation presented here: in fact, it can be regarded also as a prototypical
instance of integration over the moduli space of exotic or stringy
multi-instantons. Let us explain what we mean by this. The construction of
``brane-world'' models in which four-dimensional gauge and matter theories live
on the world-volume of suitable D-brane stacks has assumed a prominent r\^ole
for possible phenomenological applications of string theory. In this context,
non-perturbative contributions to the effective action for the gauge/matter
degrees of freedom can arise from instantonic branes, that is from branes that
are point-like in the four non-compact space-time directions. Instantonic branes
which in the internal space coincide with the D-branes that support the gauge
theory correspond to the usual gauge instanton configurations
\cite{Witten:1995gx}-\nocite{Douglas:1995bn,Green:2000ke}\cite{Billo:2002hm}.
{From} the CFT point of view, open strings suspended between instantonic and
gauge branes have four directions with mixed Neumann-Dirichlet (ND) boundary
conditions, and possess massless excitations in the Neveu-Schwarz sector
corresponding to the moduli which describe the size and gauge orientation of
field-theoretical instanton solutions.

On the other hand, instantonic branes which do not coincide with the gauge
branes in the internal directions are usually referred to as exotic or stringy
instantons. Much interest in their properties was sparkled by the realization
that they can generate terms in the effective action which are forbidden in
perturbation theory but are necessary for phenomenological applications, such as
neutrino Majorana mass terms or certain Yukawa couplings in GUT models (see
Ref.~\cite{Blumenhagen:2009qh} and references therein). {From} the CFT point of
view, mixed open strings have extra twisted directions besides the four ND
space-time directions. As a consequence, the bosonic moduli corresponding to the
size are missing and certain fermionic zero-modes become difficult to saturate.
These unwanted zero-modes must be either lifted
\cite{Blumenhagen:2007bn}-\nocite{Billo':2008pg}\cite{Billo':2008sp} or removed
by appropriate projections
\cite{Argurio:2007qk}-\nocite{Argurio:2007vqa}\cite{Bianchi:2007wy} in order to
get non-vanishing contributions.

The extension of the instanton calculus to the exotic cases is therefore of
great relevance. In some set-ups with $\cN = 1$ supersymmetry it has been shown
that novel interactions terms in the effective superpotential can arise from
sectors with a specific instanton number \cite{Blumenhagen:2009qh}. With $\cN=2$
supersymmetry, instead, one expects contributions from all sectors, in analogy
with what happens for ordinary gauge instantons in four dimensions. In this
case, in fact, using the exact Seiberg-Witten solution of $\mathcal N=2$ super
Yang-Mills (SYM) theories \cite{Seiberg:1994rs}, one can show that the effective
prepotential receives contributions from all instantons. A few years ago
\cite{Nekrasov:2002qd}, such a prediction was finally checked against the direct
evaluation of the non-perturbative effects at all instanton numbers in the
microscopic SYM theory. This remarkable computation was made possible by a
BRST-invariant reformulation of the instanton moduli action, the introduction of
suitable deformations and the use of localization techniques
\cite{Moore:1998et,Nekrasov:2002qd} \footnote{See
Ref.s~\cite{Flume:2002az}-\nocite{Bruzzo:2002xf,
Nekrasov:2003rj,Bruzzo:2003rw}\cite{Marino:2004cn} for further applications and
generalizations.}. In Ref.~\cite{Billo:2006jm}, this procedure was reproduced in
a stringy way using systems of D3/D(--1)-branes. In that context, the
localization deformations arise from interactions with a Ramond-Ramond (RR)
closed string graviphoton background.

Here we extend this approach to systems of D7/D(--1)-branes in the type
I$^\prime$ theory. This extension is not a priori obvious, given the very
different structure of the moduli space, but actually, as we will see, it
carries over in a rather natural way, and in the end it allows us to explicitly
perform the integration over the instanton moduli and check the predictions from
the heterotic string. As discussed in detail in Ref.~\cite{Billo':2009gc}, the
D7/D(--1) brane systems display the typical features of the exotic instantons in
that they have ``more than four'' ND directions (eight, in fact) and lack the
bosonic charged moduli related to the size. The gauge theory living on the
eight-dimensional world-volume of the D7-branes has a quartic action for the
gauge fields that is described by a prepotential function, analogously to the
quadratic action for the $\cN=2$ SYM theories in four dimensions. This
prepotential receives non-perturbative contributions from all numbers of
D-instantons, and here we show how to compute them relying on the BRST structure
of the instanton action, the introduction of deformations from the RR sector and
the use of localization techniques. Given the similarities of the moduli
spectra, the techniques used in this case should be useful also for the
treatment of exotic instanton contributions in four-dimensional theories.

The structure of this paper is as follows: in the next section we briefly review
the results expected from the heterotic/type I$^\prime$ duality for the
non-perturbative contributions to the quartic couplings. In Section
\ref{sec:BRSmod} we describe the BRST structure of the instanton moduli action,
which we deform by introducing a RR background in Section \ref{sec:RRback}.
Then, in Section \ref{sec:scaling} we discuss the rescalings that lead to the
localization of the moduli integrals that are explicitly evaluated in Section
\ref{sec:explicit} up to instanton number $k=5$. In the last two sections we
collect our results and present our conclusions. Finally, some technical details
on the conventions, on the interactions with the RR background and on the
evaluation of the moduli integrals are contained in three appendices.

\section{Heterotic results and duality to type I$^\prime$}
\label{sec:hd}
In order to be self-contained, we begin by briefly reviewing the heterotic
results on the quartic effective action for the system we want to consider, and
the philosophy of the stringy instanton calculus that we will apply on the type
I$^\prime$ side.
 
\subsection{Heterotic vs type I$^\prime$ results for the quartic effective
action} 
\label{subsec:hetres}
Let us consider a toroidal compactification of the SO(32) heterotic string.
Differently from what happens for the uncompactified case, the gauge quartic
terms $F^4$ and their gravitational counterparts $F^2\mathcal R^2$ and $\mathcal
R^4$ are not completely fixed by supersymmetry and anomaly cancellation, but
still are sensitive only to the BPS sector of the theory and, as such, enjoy
non-renormalization properties \cite{Bachas:1997xn}. Thus, these quartic
couplings are the natural terms to consider in order to test the duality map
between the heterotic string and the type I theory.

On the heterotic side, the quartic terms are exact at one loop and have been
computed in various toroidal compactifications with non-trivial Wilson lines.
Here we consider a compactification on a 2-torus $\mathcal T_2$ with Wilson
lines that break the gauge group SO(32) down to $\big[\mathrm{SO}(8)\big]^4$.
This case presents some interesting peculiarities since, besides the
single-trace and double-trace quartic invariants, the group SO$(8)$ possesses a
third independent invariant of order four: the Pfaffian. As a consequence, the
algebraic structure of the quartic effective action is richer. The part
containing the simple- and double-trace terms was computed in
Ref.s~\cite{Lerche:1998nx,Gutperle:1999xu,Kiritsis:2000zi}, while the Pfaffian
part was considered in Ref.~\cite{Gava:1999ky}. In our normalizations, and
denoting by $\Th$ and $\Uh$, respectively, the (complexified) K\"ahler modulus
and the complex structure of the 2-torus $\cT_2$, the quartic effective
couplings read
\begin{equation}
 \label{hr1}
 \begin{aligned}
& \frac{t_8\,\Tr F^4}{4}\,
\log\left|\frac{\eta(4\Th)}{\eta(2\Th)}\right|^4 +  \frac{t_8\,
(\Tr F^2)^2}{16}\, \log \left(\im\Th\, \im\Uh
\frac{|\eta(2\Th)|^8\, |\eta(\Uh)|^4}{|\eta(4\Th)|^4}  \right)\\
& \hspace{60pt} +2\,t_8\,\Pf F\,\log \left|\frac{\eta(\Th +
1/2)}{\eta(\Th)}\right|^4
 \end{aligned}
\end{equation}
where $\eta$ is the Dedekind function and $t_8$ is the eight-index tensor
arising in various string amplitudes \cite{Green:1987mn} (see Appendix
\ref{subapp:t8} for more details). More precisely, the notation $t_8\,\Tr F^4$
stands for
\begin{eqnarray}
t_8 \Tr& &\!\!\!\!\!\!\!\!\!F^4 \equiv  \frac{1}{2^4}
\,t_8^{\mu_1\mu_2\cdots\mu_7\mu_8} \,\Tr \big(F_{\mu_1\mu_2}\cdots
F_{\mu_7\mu_8}\big)
\label{f4}\\
& = &\Tr\Big(F_{\mu\nu}F^{\nu\rho}F^{\lambda\mu}F_{\rho\lambda} 
 +\frac{1}{2}\,F_{\mu\nu}F^{\rho\nu}F_{\rho\lambda}F^{\mu\lambda}
 -\frac{1}{4}\,F_{\mu\nu}F^{\mu\nu}F_{\rho\lambda}F^{\rho\lambda}
 -\frac{1}{8}\,F_{\mu\nu}F_{\rho\lambda}F^{\mu\nu}F^{\rho\lambda}\Big)~,
 \nonumber
\end{eqnarray}
with a similar expression for $t_8\,(\Tr F^2)^2$, while the loose 
notation $t_8\,\Pf F$ actually means
\begin{equation}
 \label{t8PfF}
t_8\,\Pf F \equiv \frac{1}{2^8}\,t_8^{\mu_1\mu_2\cdots\mu_7\mu_8}
\,\epsilon_{a_1a_2\cdots a_7a_8}\,F_{\mu_1\mu_2}^{a_1a_2}\cdots 
F_{\mu_7\mu_8}^{a_7a_8}
\end{equation}
with $a_i$'s being indices of the fundamental representation of SO(8).
It is interesting to observe that the coupling functions appearing in
front of all the three gauge-invariant structures in (\ref{hr1}) are invariant
under the modular%
\footnote{The subgroup $\Gamma_0(4)\subset \mathrm{SL}(2,\mathbb{Z})$ is
generated by $t$ and $s t^4 s$, if $t: \Th\to \Th+1$ and $s: \Th \to -1/\Th$
are the usual $\mathrm{SL}(2,\mathbb{Z})$ generators.}  subgroup $\Gamma_0(4)$
acting on $\Th$ which is preserved by the insertion of the Wilson lines. 

In this model there are also quartic interactions involving the space-time 
curvature two-form $\mathcal R$. They correspond (schematically) to the
following structures
\begin{equation}
 \label{quarticcurv}
  t_8\, \Tr F^2\, \Tr \mathcal R^2~,~~~~
  t_8\, \Tr \mathcal R^4~,~~~~
  t_8\, \big(\Tr \mathcal R^2\big)^2
\end{equation}
and, like the pure gauge terms, they are also captured exactly by a 1-loop 
heterotic computation. From the results contained for example in Ref.s
\cite{Lerche:1998nx,Kiritsis:2000zi} one can deduce that (up to an overall
convention dependent coefficient) such gravitational terms are
\begin{equation}
 \label{gravhet}
 t_8\Big[\Tr \mathcal R^4 + \frac 14 \big(\Tr \mathcal R^2\big)^2 - 16\,
\Tr F^2\, \Tr \mathcal R^2 \Big] 
\log \left(\im\Th\, \im\Uh\, |\eta(2\Th)|^4\, |\eta(\Uh)|^4\right)~. 
\end{equation}
Let us now focus on the $\Th$ dependence and introduce the parameter
\begin{equation}
 \label{qhdef}
 \qh = \ee^{2\pi\ii\, \Th}~.
\end{equation}
Then, the quartic gauge couplings (\ref{hr1}) can be rewritten as 
\begin{subequations}
 \label{exphet}
 \begin{align}
  &\,t_8\,\Tr F^4\,\Big\{\Big(\frac{\pi\ii\Th}{12}  -
    \frac 12 \sum_{k=1}^\infty \big(d_k \qh^{4k}-d_k \qh^{2k}\big)\Big) +
    \mathrm{c.c.} \Big\}   \label{expheta}\\ 
  +\,&\, t_8\,(\Tr F^2)^2\, \Big\{ \frac{1}{16}
    \log\left(\im\Th\, \im\Uh\, |\eta(\Uh)|^4\right) +
    \frac{1}{8} \,\Big( \sum_{k=1}^\infty \big(d_k \qh^{4k}-2 d_k \qh^{2k}\big)+
    \mathrm{c.c.}    \Big)  \Big\}\label{exphetb}\\
  +\,&8\,t_8\,\Pf F \, \Big(\sum_{k=1}^\infty d_{2k-1}
    \qh^{2k-1} + \mathrm{c.c}  \Big)\label{exphetc}
 \end{align}
\end{subequations}
where the coefficients $d_k$ are given by the sum of the inverse of the
divisors of $k$:
\begin{equation}
 \label{defdk}
  d_k = \sum_{\ell|k} \frac{1}{\ell}~.
\end{equation}
Likewise, the gravitational couplings (\ref{gravhet}) become
\begin{equation}
 \label{gravhet1}
\begin{aligned}
&t_8\Big[\Tr \mathcal R^4 + \frac 14 \big(\Tr \mathcal R^2\big)^2 - 16\,
\Tr F^2\, \Tr \mathcal R^2 \Big]\times\\
&\hspace{75pt}\times\Big\{\Big(\frac{\pi\ii\Th}{3}
+\,\frac{1}{2}\log\left(\im\Th\, \im\Uh\, |\eta(\Uh)|^4\right) -
2 \sum_{k=1}^\infty d_k\qh^{2k}\Big)~+~\mathrm{c.c.}\Big\}~.
\end{aligned}
\end{equation}
Written in this form, the quartic terms admit a direct interpretation in the dual type I$^\prime$
theory. To see this, let us first recall that the type I$^\prime$ theory is obtained from 
the type IIB string compactified on $\mathcal T_2$ with the orientifold projection
\begin{equation}
\Omega = \omega \,(-1)^{F_L}\,{\mathcal I}_2
\label{orientifoldparity}
\end{equation}
where $\omega$ is the world-sheet parity, $F_L$ is the left-moving world-sheet
fermion number, and ${\mathcal I}_2$ is the inversion along the two directions
of $\mathcal T_2$. The resulting theory is an unoriented string model with
sixteen supercharges. The action of $\Omega$ has four fixed-points on $\mathcal
T_2$ where four O7-planes are placed. A local cancellation of the RR tadpoles
produced by these O7-planes requires to place at each fixed-point eight
D7-branes or, equivalently, four D7 branes plus their orientifold images.
Focusing on only one of the fixed-points, we therefore have a gauge theory with
group SO(8) and $\mathcal N=2$ supersymmetry in eight dimensions.

The type I$^\prime$ model is dual to the $\big[\mathrm{SO}(8)\big]^4$ heterotic
string on $\mathcal T_2$. In particular, the duality map relates the
complexified K\"ahler modulus of the torus $\Th$ on the heterotic side and the
axion-dilaton field $\tau$ on the type I$^\prime$ side, while the complex
structure remains the same:
\begin{equation}
 \label{dualrel}
 \Th \leftrightarrow \tau \equiv C_0 + \frac{\ii}{g_s}~~,~~~~ 
 \Uh \leftrightarrow U~,
\end{equation}
where $g_s$ and $C_0$ are, respectively, the string coupling constant and the
scalar of the RR sector. Thus, on the type I$^\prime$ side, we should retrieve
exactly the results of Eq.s (\ref{exphet}) and (\ref{gravhet1}) upon the
replacement of $q_{\mathrm h}$ with
\begin{equation}
 \label{qdef}
 q \equiv \ee^{2\pi\ii\tau}~.
\end{equation}
For the single trace structure, from  Eq.~(\ref{expheta}) we expect to find a
tree-level term proportional to $\im\tau = 1/g_s$ plus a series of
non-perturbative contributions weighted by powers of $q$ which, as we will see,
are due to D-instantons. For the double trace structure  we identify in
Eq.~(\ref{exphetb}) a term proportional to $\log(\im\tau) = -\log g_s$ that
arises at 1-loop, plus a series of D-instanton contributions. The Pfaffian
structure, instead, gets only non-perturbative contributions with odd instanton
number, as we see from Eq.~(\ref{exphetc}). Finally, the quartic gravitational
couplings of type I$^\prime$ have a tree-level term proportional to $\im\tau$, a
1-loop term proportional to $\log(\im\tau)$ and a series of non-perturbative
contributions with even instanton number, as indicated in the second line of
Eq.~(\ref{gravhet1}).

In the literature, the heterotic results we described above have been compared
\cite{Lerche:1998nx,Kiritsis:2000zi} with F-theory compactified on K3, which has
been argued \cite{Sen:1996vd} to represent a geometrized non-perturbative
version of the type I$^\prime$ model. Our aim is instead to compare them with a
direct computation of non-perturbative D-instanton effects in the type
I$^\prime$ string theory. The general philosophy behind such a computation is
briefly summarized in the next subsection.

\subsection{D-instanton contributions to the quartic effective action in type
I$^\prime$} 
\label{Dcont}

When $k$ D(--1)-branes are added to the D7-branes, new open string sectors
appear, corresponding to open strings with at least one endpoint attached to the
D(--1)'s. The excitations of such strings carry no momentum; we call them
moduli%
\footnote{If one considers systems of D3/D(--1)-branes (or more
generally D$(p+4)$/D$p$-branes), the moduli excitations in the $k$ D-instanton
sectors are in full correspondence with the moduli of the classical instanton
solutions with instanton number $k$, as encoded in the ADHM construction. The
name ``moduli'' continues to be used in more general situations where the
correspondence with classical solutions might be less immediate.}, and
collectively denote them as $\cM_{(k)}$. The action for $k$ D-instantons has a
classical part, $\mathcal S_{\mathrm{cl}}=-2\pi\ii\tau\, k$, and a
moduli-dependent part, $\mathcal S(\cM_{(k)})$, arising from disk diagrams with
at least a portion of their boundary attached to the D(--1)'s. There are also
mixed disk diagrams describing the interactions between the moduli and the gauge
fields living on the D7-branes which are encoded in a (chiral) superfield
$\Phi$. By including also these diagrams, the moduli action $\mathcal
S(\cM_{(k)})$ is promoted to $\mathcal S(\cM_{(k)},\Phi)$.

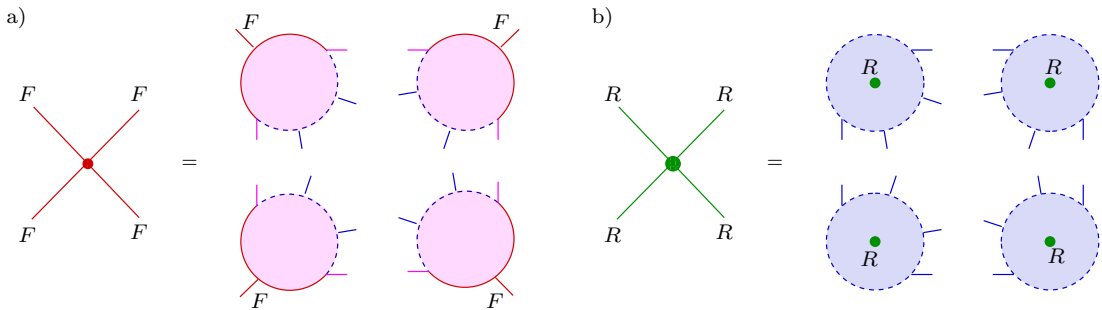
\begin{figure}[hbt]
 \begin{center}
 \begin{picture}(0,0)%
\includegraphics{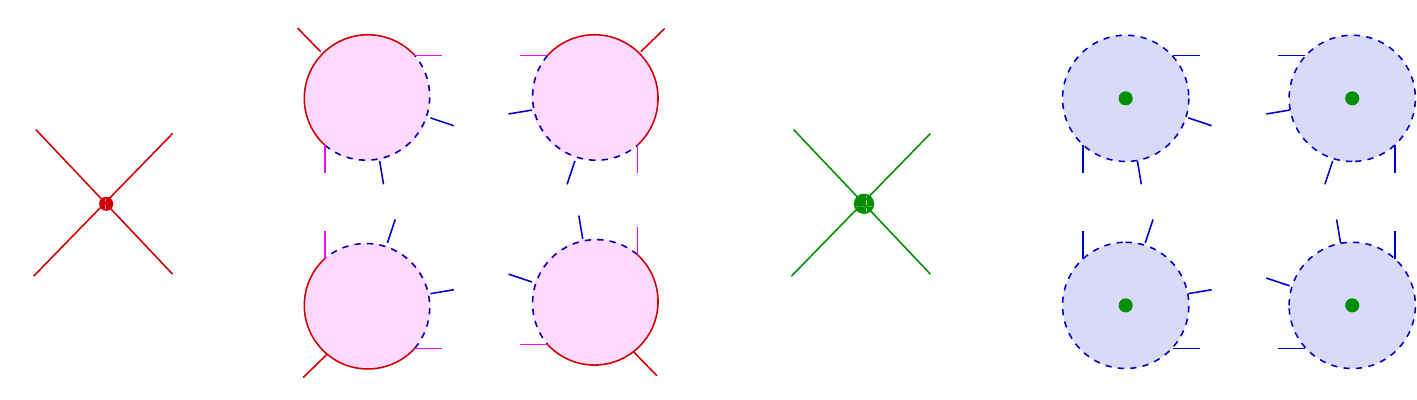}%
\end{picture}%
\setlength{\unitlength}{987sp}%
\begingroup\makeatletter\ifx\SetFigFontNFSS\undefined%
\gdef\SetFigFontNFSS#1#2#3#4#5{%
  \reset@font\fontsize{#1}{#2pt}%
  \fontfamily{#3}\fontseries{#4}\fontshape{#5}%
  \selectfont}%
\fi\endgroup%
\begin{picture}(27200,7657)(-14,-6806)
\put(4351,-3136){\makebox(0,0)[lb]{\smash{{\SetFigFontNFSS{7}{8.4}{\familydefault}{\mddefault}{\updefault}=}}}}
\put(301,-1486){\makebox(0,0)[lb]{\smash{{\SetFigFontNFSS{8}{9.6}{\familydefault}{\mddefault}{\updefault}$F$}}}}
\put(3076,-4936){\makebox(0,0)[lb]{\smash{{\SetFigFontNFSS{8}{9.6}{\familydefault}{\mddefault}{\updefault}$F$}}}}
\put(3076,-1486){\makebox(0,0)[lb]{\smash{{\SetFigFontNFSS{8}{9.6}{\familydefault}{\mddefault}{\updefault}$F$}}}}
\put(301,-5011){\makebox(0,0)[lb]{\smash{{\SetFigFontNFSS{8}{9.6}{\familydefault}{\mddefault}{\updefault}$F$}}}}
\put(  1,464){\makebox(0,0)[lb]{\smash{{\SetFigFontNFSS{8}{9.6}{\familydefault}{\mddefault}{\updefault}a)}}}}
\put(12076,314){\makebox(0,0)[lb]{\smash{{\SetFigFontNFSS{8}{9.6}{\familydefault}{\mddefault}{\updefault}$F$}}}}
\put(5851,314){\makebox(0,0)[lb]{\smash{{\SetFigFontNFSS{8}{9.6}{\familydefault}{\mddefault}{\updefault}$F$}}}}
\put(11926,-6661){\makebox(0,0)[lb]{\smash{{\SetFigFontNFSS{8}{9.6}{\familydefault}{\mddefault}{\updefault}$F$}}}}
\put(6076,-6661){\makebox(0,0)[lb]{\smash{{\SetFigFontNFSS{8}{9.6}{\familydefault}{\mddefault}{\updefault}$F$}}}}
\put(18901,-3136){\makebox(0,0)[lb]{\smash{{\SetFigFontNFSS{7}{8.4}{\familydefault}{\mddefault}{\updefault}=}}}}
\put(14851,-1486){\makebox(0,0)[lb]{\smash{{\SetFigFontNFSS{8}{9.6}{\familydefault}{\mddefault}{\updefault}$R$}}}}
\put(17626,-4936){\makebox(0,0)[lb]{\smash{{\SetFigFontNFSS{8}{9.6}{\familydefault}{\mddefault}{\updefault}$R$}}}}
\put(17626,-1486){\makebox(0,0)[lb]{\smash{{\SetFigFontNFSS{8}{9.6}{\familydefault}{\mddefault}{\updefault}$R$}}}}
\put(14851,-5011){\makebox(0,0)[lb]{\smash{{\SetFigFontNFSS{8}{9.6}{\familydefault}{\mddefault}{\updefault}$R$}}}}
\put(25801,-811){\makebox(0,0)[lb]{\smash{{\SetFigFontNFSS{8}{9.6}{\familydefault}{\mddefault}{\updefault}$R$}}}}
\put(25876,-5536){\makebox(0,0)[lb]{\smash{{\SetFigFontNFSS{8}{9.6}{\familydefault}{\mddefault}{\updefault}$R$}}}}
\put(21226,-5611){\makebox(0,0)[lb]{\smash{{\SetFigFontNFSS{8}{9.6}{\familydefault}{\mddefault}{\updefault}$R$}}}}
\put(21226,-811){\makebox(0,0)[lb]{\smash{{\SetFigFontNFSS{8}{9.6}{\familydefault}{\mddefault}{\updefault}$R$}}}}
\put(14551,464){\makebox(0,0)[lb]{\smash{{\SetFigFontNFSS{8}{9.6}{\familydefault}{\mddefault}{\updefault}b)}}}}
\end{picture}%
 \end{center}
 \caption{a) A quartic interaction vertex for the gauge field $F$ can be
 induced by mixed disks having part of their boundary attached to the
 D-instantons and carrying the insertion of a vertex for $F$ and of moduli
 vertices. The above diagram is connected by the integration over the moduli.
 b) Instanton disks with an insertion of a closed vertex can produce
 curvature interactions through the moduli integration.}
 \label{fig0}
\end{figure}

Non-perturbative contributions to the effective action for $\Phi$
arise upon integration over the moduli of the exponentiated field-dependent
action of the D-instantons \cite{Green:2000ke,Billo:2002hm}:
\begin{equation}
 \label{itoe}
\sum_k \ee^{-\cS_{\mathrm{cl}}}\int d\cM_{(k)} \, \ee^{-\cS(\cM_{(k)},\Phi)} =
 \sum_k q^{k} \int d\cM_{(k)} \, \ee^{-\cS(\cM_{(k)},\Phi)}~.
\end{equation}
In particular, the integration over the moduli of those mixed disks that are
sources for some components of $\Phi$, can produce new effective couplings, as
represented in Fig.~\ref{fig0}a for the quartic interaction among four gauge
field strengths $F$.

As already noted in the literature (see for instance
Ref.s~\cite{Bachas:1997mc,Kiritsis:1997hf,Bachas:1997xn,Lerche:1998nx,
Gutperle:1999xu,Kiritsis:2000zi}), these D-instanton induced couplings have
potentially the right structure to agree with the heterotic results reported in
\eq{exphet}. Indeed, the sum over the number $k$ of D-instantons is weighted by
$q^k$ and the dimensionality of the moduli measure $d\cM_{(k)}$ implies that the
effective contributions must be quartic in the gauge fields for all $k$
\cite{Billo':2009gc}.

To turn the schematic expression (\ref{itoe}) into a real computational tool, it
is necessary to precisely identify the moduli $\cM_{(k)}$, compute their
field-dependent action $\cS(\cM_{(k)},\Phi)$ and explicitly perform the matrix
integrals. The latter task is far from being trivial. The 2-instanton case was
already considered in Ref.~\cite{Gutperle:1999xu} where it was argued that the
correct gauge-invariant structures $\Tr F^4$ and $(\Tr F^2)^2$ are obtained from
the integration over the moduli $\cM_{(2)}$ with a relative coefficient in
agreement with \eq{exphet}. However, to reach more solid conclusions, an
analysis at higher values of $k$ is necessary. 

In this case, the only viable route to get explicit results is a generalization
of the methods that were successfully applied for the instanton calculus in
$\mathcal N=2$ SYM theories in four dimensions. This requires to exploit the
particular algebraic structure and the supersymmetry of the moduli action
$\cS(\cM_{(k)},\Phi)$ and write it as a $Q$-exact expression with respect to a
suitable BRST charge $Q$, in such a way that the localization techniques
\cite{Moore:1998et,Nekrasov:2002qd} can be applied. These involve the
introduction of deformations of $\cS(\cM_{(k)},\Phi)$ which, while not altering
the final result, may drastically simplify the computation. The needed
deformations, which could be introduced {\it ad hoc} from a purely mathematical
point of view, arise naturally from mixed disk diagrams describing the
interaction of the moduli $\cM_{(k)}$ with closed string graviphoton backgrounds
from the RR sector of the theory. Treating the RR field-strengths as constant
parameters to be put to zero at the end of the computation allows to write
explicit contour integral expressions which, in principle, can be evaluated for
any $k$, and from which the quartic effective action for the gauge fields can be
extracted. If we consider the RR fields as genuine, dynamical graviphotons
sitting in the same supermultiplet $\cW$ of the curvature two-form $\mathcal R$,
we can generalize \eq{itoe} and use a field-dependent moduli action
$\cS(\cM_{(k)},\Phi,\cW)$ that contains also gravitational terms. Then, the
corresponding D-instanton partition functions will yield also the $\Tr F^2\, \Tr
\mathcal R^2$ and $\Tr \mathcal R^4$ interactions (see for example
Fig.~\ref{fig0}b) which from the heterotic side are given in \eq{gravhet}.

This procedure will be described in great detail in the following sections.

\section{The D7/D(--1) system and its BRST structure}
\label{sec:BRSmod}

We now discuss the main features of the D7/D(--1) system in the type I$^\prime$
theory, both at the perturbative and the non-perturbative level.

\subsection{The perturbative sectors}
\label{subsec:gauge}
As we have already explained, the world-volume theory on the eight D7-branes
located at one of the orientifold fixed points of the type I$^\prime$ string
model is an eight-dimensional gauge theory with sixteen supercharges and gauge
group $\mathrm{SO}(8)$. Its bosonic action contains, besides the usual
Yang-Mills term, also terms of higher order in the field strength $F$ and its
covariant derivatives. Among them, a crucial r\^ole for our purposes is played
by the conformally invariant tree-level quartic terms
\begin{equation}
S_{(4)} = -\frac{1}{96\pi^3 g_s}\int \!d^8x \,\,t_8\mathrm{Tr}
F^4 - \frac{\ii\,C_0}{192\pi^3}\int \Tr\big(F\wedge
F\wedge F\wedge F\big)~.
\label{s4}
\end{equation} 
Introducing the chiral superfield
\begin{equation}
 \Phi(x,\theta) = \phi(x) + \sqrt{2}\,\theta\Lambda(x) +\frac{1}{2}\,
\theta\gamma^{\mu\nu}\theta\,F_{\mu\nu}(x) + \ldots~,
\label{Phi}
\end{equation}
where $\Lambda$ is the gaugino and $\phi$ a complex scalar, 
the quartic action (\ref{s4}) can be conveniently rewritten as
\begin{equation}
S_{(4)} = \frac{1}{(2\pi)^4}\int \!d^8x
\,d^8\theta~\Big[\frac{\ii\pi\tau}{12}\,\Tr \Phi^4\Big]+~\mbox{c.c.}~,
\label{s41}
\end{equation} 
where $\tau$ is the axion-dilaton combination appearing in \eq{dualrel}.

Other quartic terms are produced at 1-loop. Indeed, as shown for example in
Section 4.2 of Ref.~\cite{Billo':2009gc}, the annulus and M\"obius diagrams for
this brane system yield the following (divergent) contribution%
\footnote{Note that for eight D7-branes there is no contribution to
$\Tr F^4$ at 1-loop.}
\begin{equation}
 \label{1loop}
-\frac{1}{128\pi^4}\int \!d^8x \,\,t_8\big(\Tr F^2 \big)^2
\Big[\int_0^\infty \frac{dt}{2t}\,\Gamma(t) \Big]
\end{equation}
where $\Gamma(t)$ represents the sum over the winding modes in the two compact 
transverse directions, given by
\begin{equation}
 \Gamma(t) = \sum_{(r_1,r_2)\in \mathbb Z^2} \ee^{-2\pi t\,
\frac{|r_1+r_2U|^2 \mathrm{Im}T}{\mathrm{Im}U}}
\label{W}
\end{equation}
with $U$ and $T$ being, respectively, the complex and K\"ahler structures of the
2-torus $\mathcal T_2$. The integral over the modular loop parameter $t$ can be
computed using the regularization procedure introduced in
Ref.~\cite{Dixon:1990pc} (and reviewed for example in Appendix A of
Ref.~\cite{Billo:2007sw}) with the result
  \begin{equation}
   \label{wt}
  \begin{aligned}
   \int_0^\infty \frac{dt}{2t}\,\Gamma(t) & =-\frac{1}{2}\,
   \log\big(\alpha'\mu^2\big)
   -\frac{1}{2}\,\log\Big(\frac{\mathrm{Im}U\,|\eta(U)|^4}{\mathrm{Im}T}\Big)\\
   & = -\frac{1}{2}\,\log\Big(\frac{\mu^2}{M_{\mathrm{P}}^2}\Big) + 
   \Delta^{\mathrm{1-loop}}~.
  \end{aligned}
  \end{equation}
Here $\mu$ is a low-energy scale that regularizes the IR divergence due to the
massless open string states circulating in the loop, while $M_{\mathrm{P}}$ is
the eight-dimensional Planck mass 
 \begin{equation}
  \label{mplanck}
 M_{\mathrm{P}}^2 = \frac{\mathrm{Im}T}{\alpha'\,g_s}
 \end{equation}
 which serves as UV cut-off in the field theory. 
Finally, $\Delta^{\mathrm{1-loop}}$ represents the (finite) threshold
corrections given by
\begin{equation}
 \label{threshold}
\Delta^{\mathrm{1-loop}} = -\frac{1}{2}\,\log\big(\mathrm{Im}\tau\,
\mathrm{Im}U\,|\eta(U)|^4\big)~.
\end{equation}
{From} these results, we therefore find the following 1-loop term in the 
effective action
\begin{equation}
 \label{squartic1loop}
\begin{aligned}
 S_{(4)}^{\mathrm{1-loop}} &=\frac{1}{256\pi^4}\int \!d^8x \,\,
\log\big(\mathrm{Im}\tau\,\mathrm{Im}U\,|\eta(U)|^4\big)
\,t_8\big(\Tr F^2 \big)^2\\
&= \frac{1}{(2\pi)^4}\int \!d^8x \,d^8\theta~\Big[\frac{1}{32}\,
\log\big(\mathrm{Im}\tau\,\mathrm{Im}U\,|\eta(U)|^4\big)\big(\Tr
\Phi^2\big)^2\Big]+~\mbox{c.c.}
\end{aligned}
\end{equation}
which has to be added to the tree-level contribution (\ref{s41}). Due to
$\mathcal N=2$ supersymmetry, there are no higher-loop quartic terms in the
effective action.

\subsection{The non-perturbative sectors}
As discussed in Ref.~\cite{Billo':2009gc}, the non-perturbative sectors of this
theory can be described by adding $k$ D(--1)-branes in the same fixed point
where the D7's are located. The D(--1)-branes are sources for the RR scalar
$C_0$; thus, considering the Wess-Zumino part of the D7 action (\ref{s4}), it
follows that $k$ D-instantons correspond to a gauge field configuration with
fourth Chern number
\begin{equation}
 c_{(4)} = \frac{1}{4!\,(2\pi)^4}\,\int \Tr\big(F\wedge
F\wedge F\wedge F\big) = k~.
\label{c4} 
\end{equation}
Moreover, this gauge field configuration must be such that its classical quartic
action reduces to $k$ times the D-instanton action \cite{Billo':2009gc}, {\it
i.e.} $S_{(4)} = -{2\pi\ii}\,\tau\, k$.

The physical excitations of the open strings with at least one end-point on the
D-instantons account for the moduli $\cM_{(k)}$ of such instanton-like
configurations. The neutral sector, corresponding to D(--1)/D(--1) open strings,
comprises the moduli that do not transform under the gauge group and includes,
in an ADHM inspired notation, the vector $a_\mu$ and the scalar $\chi$ (plus its
conjugate $\bar\chi$) in the Neveu-Schwarz sector, and the chiral and
anti-chiral fermions $M^\alpha$ and $\lambda_{\dot\alpha}$ in the Ramond sector.
The bosonic moduli have canonical dimensions of (length)$^{-1}$, while the
fermionic ones have canonical dimensions of (length)$^{-\frac{3}{2}}$. 

All these neutral moduli are $k\times k$ matrices, but the consistency with the
orientifold projection on the D7-branes requires that $\chi$, $\bar\chi$ and
$\lambda_{\dot\alpha}$ transform in the anti-symmetric (or adjoint)
representation of $\mathrm{SO}(k)$, while $a_\mu$ and $M^\alpha$ must be in the
symmetric one. The diagonal parts of $a_\mu$ and $M^\alpha$ represent the
bosonic and fermionic Goldstone modes of the (super)translations of the
D7-branes world-volume that are broken by the D-instantons and thus can be
identified with the bosonic and fermionic coordinates $x_\mu$ and
$\theta^\alpha$ of the eight-dimensional superspace. More precisely, we have
\begin{equation}
 x_\mu= (2\pi\alpha')\, \mathrm{tr}\big(a_\mu\big)\quad,\quad
\theta^\alpha = (2\pi\alpha')\, \mathrm{tr}\big(M^\alpha\big)~,
\label{xtheta}
\end{equation}
where the factors of $\alpha'$ have been introduced to give $x_\mu$ and
$\theta^\alpha$ the appropriate dimensions.

The open strings stretching between the D-instantons and the D7-branes account
for the charged moduli, which transform in the fundamental representations of
both $\mathrm{SO}(8)$ and $\mathrm{SO}(k)$. The D7/D(--1) open strings have
eight ND directions and thus, as discussed for example in Ref.
\cite{Billo':2009gc}, it is not possible to find bosonic excitations that
satisfy the physicity conditions. The absence of charged bosonic moduli is the
hallmark of the ``exotic'' instanton configurations, and has to be contrasted
with what happens in the D3/D(--1) systems where, instead, physical bosonic
moduli, related to the gauge instanton size, exist. On the other hand, the
fermionic Ramond sector of the D7/D(--1) system is not empty and contains
physical moduli, denoted as $\mu$ and $\bar\mu$ depending on the orientation.
They are, respectively, $k\times N$ and $N\times k$ matrices (with $N=8$ in our
specific case). Since the orientifold parity (\ref{orientifoldparity}) exchanges
the two orientations, in the Type I$^\prime$ theory $\mu$ and $\bar\mu$ are not
independent of each other but are related according to $\bar\mu = -
\,{}^{\mathrm{t}}\!\mu$.

For all the physical moduli $\cM_{(k)}$ listed above, it is possible to write
vertex operators of conformal dimension 1 and use them to obtain the moduli
action $\cS(\cM_{(k)})\equiv \cS$ by computing disk amplitudes along the lines
discussed in Refs.~\cite{Green:2000ke,Billo:2002hm,Billo:2006jm}. As a result
one finds \cite{Billo':2009gc}
\begin{equation}
{\mathcal S}=
{\mathcal S}_{\mathrm{cubic}}+{\mathcal S}_{\mathrm{quartic}}
+{\mathcal S}_{\mathrm{mixed}}
\label{smod}
\end{equation}
where%
\footnote{Here we use slightly different conventions for the $\mu$'s as
compared to Ref.~\cite{Billo':2009gc}.}
\begin{subequations}
\begin{align}
\label{cubic}
& {\mathcal S}_{\mathrm{cubic}} = 
\frac{1}{g_0^2}\,\tr \Big\{ \ii\,\lambda_{\dot\alpha}
\gamma_\mu^{\dot\alpha\beta} [a^\mu,M_\beta] -
\frac{\ii}{\sqrt{2}}\,\lambda_{\dot\alpha}[\chi,\lambda^{\dot\alpha}] 
- \frac{\ii}{\sqrt{2}}\,M^\alpha[\bar\chi,M_\alpha] \Big\}~,\\
\label{quartic}
& {\mathcal S}_{\mathrm{quartic}}=
\frac{1}{g_{0}^2}\, \tr
\Big\{\!-\frac{1}{4} \left[ a_\mu, a_\nu \right]^2 -
\comm{a_\mu}{\bar\chi }\comm{a^\mu}{\chi}  + \frac12
\comm{\bar\chi}{\chi}^2 \Big\}~,\\
\label{mixed}
& {\mathcal S}_{\mathrm{mixed}} = 
\frac{1}{g_0^2} \tr \Big\{\!-\ii\sqrt{2}\,
{}^{\mathrm{t}}\!\mu\,\chi\, \mu \Big\}~,
\end{align}
\end{subequations}
with $g_0$ being the Yang-Mills coupling constant in zero dimensions:
\begin{equation}
g_0^2 = \frac{g_s}{4\pi^3\alpha'^2}~.
\label{g0}
\end{equation}
Indeed, the total action (\ref{smod}) can also be derived by dimensionally
reducing the $\mathcal N=1$ supersymmetric Yang-Mills theory with fundamental
matter from ten to zero dimensions.

The quartic interactions $\left[ a_\mu, a_\nu \right]^2 $ appearing in
(\ref{quartic}) can be disentangled by introducing seven auxiliary fields $D_m$
($m=1,\ldots,7$) and replacing ${\mathcal S}_{\rm quartic}$ with
\begin{equation}
\label{quartic1}
{\mathcal S}'_{\mathrm{quartic}} = \frac{1}{g_{0}^2}\, {\rm tr}\,
\Big\{
\frac{1}{2}D_m D^m - \frac{1}{2}D_m (\tau^m)_{\mu\nu} \comm{a^{\mu}}{a^{\nu}}
- \comm{a_\mu}{\bar\chi }\comm{a^\mu}{\chi}  + 
\frac{1}{2}\comm{\bar\chi}{\chi}^2
\Big\}~.
\end{equation}
Here $(\tau^m)_{\mu\nu}$ are the $\gamma$-matrices of $\mathrm{SO}(7)$ (related
to the octonionic structure constants as shown in Appendix \ref{appa}) implying
that the eight-dimensional indices $\mu,\nu,\ldots$ are interpreted as spinorial
indices of SO$(7)$. The resulting moduli action is similar in structure to the
one considered in Ref.~\cite{Moore:1998et} in the analysis of the so-called
Yang-Mills integrals in $d=10$. By eliminating $D_m$ through the field equation 
\begin{equation}
 D_m = \frac{1}{2}(\tau^m)_{\mu\nu} \comm{a^{\mu}}{a^{\nu}}~,
\end{equation}
and by exploiting the properties of the $\tau^m$ matrices, one can easily see
that ${\mathcal S}'_{\mathrm{quartic}}$ is equivalent to the initial action
(\ref{quartic}).

Let us now reorganize the moduli in an ``octonionic'' form ({\it i.e.} in
representations of $\mathrm{SO}(7)$) by relabeling some of them as follows:
\begin{equation}
 \label{octmod}
M_\alpha  \to M_\mu \equiv (M_m, -M_8)~,\quad
\lambda_{\dot\alpha}  \to (\lambda_m,\eta) \equiv (\lambda_m,\lambda_8)~.
\end{equation}
In other words, the chiral moduli $M^\alpha$ are assembled into a spinor of
$\mathrm{SO}(7)$, while the anti-chiral moduli $\lambda_{\dot\alpha}$ are split
into a vector and a scalar of $\mathrm{SO}(7)$. Then, by using the explicit form
of the $\gamma^\mu$ matrices given in Appendix \ref{appa}, we can rewrite the
cubic action (\ref{cubic}) as
\begin{equation}
\begin{aligned}
{\mathcal S}'_{\mathrm{cubic}} = &\,
\frac{1}{g_{0}^2}\, \tr
\Big\{\lambda_m (\tau^m)_{\mu\nu}\comm{a^\mu}{M^\nu} + \eta \comm{a_\mu}{M^\mu} 
-\frac{\ii}{\sqrt{2}}\, M_{\mu} \comm{\bar\chi}{M^\mu} \\
&~~~~~~~~~ -\frac{\ii}{\sqrt{2}} \,\eta \comm{\chi}{\eta} - 
\frac{\ii}{\sqrt{2}} \,\lambda_m
\comm{\chi}{\lambda^m}\Big\}~.
\end{aligned}
\label{cubic1}
\end{equation}
It is also convenient to replace the mixed action (\ref{mixed}) with
\begin{equation}
{\mathcal S}'_{\mathrm{mixed}} = 
\frac{1}{g_0^2} \tr \Big\{\!{}^{\mathrm{t}}w\,w-
\ii\sqrt{2}\,{}^{\mathrm{t}}\!\mu\,\chi\, \mu \Big\}
\label{mixed1}
\end{equation}
where $w$ is an auxiliary field in the fundamental representations of
$\mathrm{SO}(k)$ and $\mathrm{SO}(8)$ which does not interact with any other
modulus. Even if this auxiliary field looks trivial, it is nevertheless useful
to introduce it for reasons which will become clear in a moment%
\footnote{We remark that just like the physical moduli, also the auxiliary
fields, including $w$, can be given an explicit string description in terms of
vertex operators with conformal dimension 1, see Appendix \ref{appb} and
Refs.~\cite{Billo:2002hm,Billo:2006jm} for details.}.

The total action 
\begin{equation}
{\mathcal S}'=
{\mathcal S}'_{\mathrm{cubic}}+{\mathcal S}'_{\mathrm{quartic}}
+{\mathcal S}'_{\mathrm{mixed}}
\label{smod1}
\end{equation}
is invariant under transformations of the D-instanton group $\mathrm{SO}(k)$, of
the gauge group $\mathrm{SO}(8)$ and of the auxiliary group $\mathrm{SO}(7)$. It
is also invariant under the following fermionic BRST transformations
\begin{equation}
 \label{Q}
\begin{aligned}
& Q a^\mu  = M^\mu~,~~~ Q M^\mu = \ii\sqrt{2}\, \comm{\chi}{a^\mu}~, \\
& Q\lambda_m = D_m~,~~~ Q D_m  =  \ii\sqrt{2}\, \comm{\chi}{\lambda_m}~,\\
& Q \bar\chi = -\ii \sqrt{2} \eta~,~~~ 
Q\eta  = -\comm{\chi}{\bar\chi}~,~~~ Q \chi = 0~, \\
& Q \mu = w~,~~~ Q w = \ii\sqrt{2}\,\chi\,\mu~.
\end{aligned}
\end{equation}
The BRST charge $Q$ is one of the supersymmetries that are preserved both by the
D-instantons and by the D7-branes; more precisely, after using (\ref{octmod}),
one can see that $Q$ is the component of the anti-chiral supercharge
$Q_{\dot\alpha}$ corresponding to $\dot\alpha=8$ (see \eq{qal}). The BRST charge
is nilpotent up to an (infinitesimal) $\mathrm{SO}(k)$ rotation parameterized by
$\ii\sqrt{2}\,\chi$. Indeed, on the moduli transforming either in the symmetric
or in the anti-symmetric representation of $\mathrm{SO}(k)$, such as $a^\mu$ or
$\lambda_m$ respectively, we have%
\footnote{Independently of its symmetry properties, any $k\times k$ matrix
$M^{IJ}$ transforms under an $\mathrm{SO}(k)$ rotation $R$ as
$M^{IJ} \to R^I_{~K} R^J_{~L} M^{KL} = R^I_{~K}M^{KL} ({}^tR)_L^{~J} = (R M
R^{-1})^{IJ}$.
If $R=\exp(A)$, with $A$ an antisymmetric matrix whose elements parameterize
the rotation, to first order we have
$\delta M = \comm{A}{M}$.
}
\begin{equation}
 Q^2\,\bullet = \ii\sqrt{2}\,\comm{\chi}{\bullet}~,
\label{Q2}
\end{equation}
while on the moduli transforming in the fundamental representation of
$\mathrm{SO}(k)$, like $\mu$ or $w$, we have
\begin{equation}
 Q^2\,\bullet = \ii\sqrt{2}\,\chi\,\bullet~.
\label{Q2f}
\end{equation}
The two BRST actions (\ref{Q2}) and (\ref{Q2f}) can be combined into a single 
formula by writing
\begin{equation}
 Q^2\,\bullet = T_{\mathrm{SO}(k)}\big(\ii\sqrt{2}\,\chi\big)\bullet
\label{Q2t}
\end{equation}
where $T_{\mathrm{SO}(k)}\big(\ii\sqrt{2}\chi\big)$ denotes an infinitesimal 
rotation of SO$(k)$, parameterized by $\ii\sqrt{2}\chi$, in the appropriate
representation of the modulus on which it acts.

By exploiting the above properties and using the gauge-invariance under
$\mathrm{SO}(k)$, one can easily show that the total moduli action (\ref{smod1})
is $Q$-exact; indeed
\begin{equation}
 {\mathcal S}' = Q\,\Xi
\label{SQXI}
\end{equation}
with the ``gauge fermion'' given by
\begin{equation}
 \label{gaugefermion}
\Xi = \frac{1}{g_0^2} \,\tr \Bigl\{ \!
\frac{1}{2} D_m \lambda^m - \frac12\lambda_m (\tau^m)_{\mu\nu} 
\comm{a^\mu}{a^\nu} + \frac{\ii}{\sqrt{2}}
\bar\chi \comm{a_\mu}{M^\mu} - \frac{1}{2}\eta \comm{\chi}{\bar\chi}
+{}^{\mathrm{t}}\!\mu\,w
\Bigr\}~.
\end{equation}
This property will play a crucial r\^ole in discussing the localization of the
integral on the instanton moduli space, as we will see in Section 
\ref{sec:scaling}. 

Let us now discuss the interactions among the instanton moduli and the gauge
fields propagating on the world-volume of the D7-branes, which we have combined
into the superfield (\ref{Phi}). Such interactions can be easily obtained by
computing mixed disk amplitudes involving both vertex operators for moduli and
vertex operators for dynamical fields, as discussed in detail in
Refs.~\cite{Billo:2002hm,Billo:2006jm} for the analogous D(--1)/D3 systems. In
the present case the result is
\begin{equation}
 \label{intPhi}
\frac{1}{g_0^2}\tr \Big\{\ii\sqrt{2}\,{}^{\mathrm{t}}\!\mu\,\mu\,
\Phi(x,\theta)\Big\}
\end{equation}
which has to be added to the moduli action (\ref{smod1}). For our later purposes
it is enough to focus on the dependence on the vacuum expectation value
\begin{equation}
 \label{vev}
\phi = \langle \Phi(x,\theta) \rangle ~,
\end{equation}
and hence we will consider the following modified mixed action
\begin{equation}
 {\mathcal S}'_{\mathrm{mixed}}(\phi) = {\mathcal S}'_{\mathrm{mixed}} + 
\frac{1}{g_0^2} \tr \Big\{\ii\sqrt{2}\,{}^{\mathrm{t}}\!\mu\,\mu\,\phi\Big\}~.
\label{smixed2}
\end{equation}
Then the total moduli action becomes
\begin{equation}
{\mathcal S}'(\phi) = {\mathcal S}'_{\mathrm{cubic}}+ 
{\mathcal S}'_{\mathrm{quartic}} + {\mathcal S}'_{\mathrm{mixed}}(\phi)~.
 \label{smod2}
\end{equation}
It is not difficult to realize that the above $\phi$-dependent terms can be
obtained by deforming the action of the BRST charge $Q$ on the auxiliary field
$w$ and replacing the last equation of (\ref{Q}) by
\begin{equation}
 \label{Qw}
Q w = \ii\sqrt{2}\,\chi\,\mu - \ii\sqrt{2}\,\mu\,\phi
\end{equation}
with all the rest, including the gauge fermion (\ref{gaugefermion}), unchanged.
Notice that with the deformation (\ref{Qw}) the BRST charge becomes nilpotent
not only up to infinitesimal rotations of $\mathrm{SO}(k)$, but also up to
infinitesimal rotations of the gauge group $\mathrm{SO}(8)$, parameterized
respectively by $\ii\sqrt{2}\,\chi$ and $-\ii\sqrt{2}\,\phi$. Thus, Eq.
(\ref{Q2t}) gets replaced by
\begin{equation}
 Q^2\,\bullet = 
 T_{\mathrm{SO}(k)}\big(\ii\sqrt{2}\,\chi\big)\bullet
-\,T_{\mathrm{SO}(8)}\big(\ii\sqrt{2}\,\phi\big)\bullet~.
\label{Q2tt}
\end{equation}
Clearly, $T_{\mathrm{SO}(8)}\big(\ii\sqrt{2}\,\phi\big)$ is non-trivial only on
$\mu$ and $w$, which are the only charged moduli transforming under the gauge
group $\mathrm{SO}(8)$. Finally, using (\ref{Q2tt}) one can easily show that
\begin{equation}
 {\mathcal S}'_{\mathrm{mixed}}(\phi) = \frac{1}{g_0^2}\,
\tr \Big\{{}^{\mathrm{t}}w\,w 
+ {}^{\mathrm{t}}\!\mu\,Q^2\,\mu\Big\}~,
\label{mixed3}
\end{equation}
where $Q^2$ is represented as an $(8k\times 8k)$ matrix acting in the tensor
product of the vector representations of $\mathrm{SO}(k)$ and $\mathrm{SO}(8)$
that are the representations under which the $\mu$'s transform.

We conclude our description of the D7/D(--1) system of Type I$^\prime$ by
summarizing in Table 1 the transformation properties of the various moduli under
$\mathrm{SO}(k)$, $\mathrm{SO}(8)$ and $\mathrm{SO}(7)$, as well as their
scaling dimensions.

\begin{table}[htb]
\label{tab:10}
\begin{center}
 \begin{tabular}{|c|c|c|c|c|}
\hline
 {\phantom{\vdots}}&$\mathrm{SO}(k)$&
 $\mathrm{SO}(8)$ & $\mathrm{SO}(7)$ & dimensions \\
\hline
${\phantom{\vdots}}a^\mu$&$\mathrm{symm}$ & $\mathbf{1}$ & $\mathbf{8}_s$ & (length)$^{-1}$ \\
${\phantom{\vdots}}M^\mu$&$\mathrm{symm}$ & $\mathbf{1}$ & $\mathbf{8}_s$ & (length)$^{-3/2}$ \\
${\phantom{\vdots}}D_m$&$\mathrm{adj}$ & $\mathbf{1}$ & $\mathbf{7}$ & (length)$^{-2}$ \\
${\phantom{\vdots}}\lambda_m$&$\mathrm{adj}$ & $\mathbf{1}$ & $\mathbf{7}$ & (length)$^{-3/2}$ \\
${\phantom{\vdots}}\bar\chi$&$\mathrm{adj}$ & $\mathbf{1}$ & $\mathbf{1}$ & (length)$^{-1}$ \\
${\phantom{\vdots}}\eta$&$\mathrm{adj}$ & $\mathbf{1}$ & $\mathbf{1}$ & (length)$^{-3/2}$ \\
${\phantom{\vdots}}\chi$&$\mathrm{adj}$ & $\mathbf{1}$ & $\mathbf{1}$ & (length)$^{-1}$ \\
${\phantom{\vdots}}\mu$&$\mathbf{k}$ & $\mathbf{8}_v$ & $\mathbf{1}$ & (length)$^{-3/2}$ \\
${\phantom{\vdots}}w$&$\mathbf{k}$ & $\mathbf{8}_v$ & $\mathbf{1}$ & (length)$^{-2}$ \\
\hline
\end{tabular}
\caption{Transformation properties and scaling dimensions of the moduli in the 
D(--1)/D7 system.}
\end{center}
\end{table}

\section{Deformation by a RR background}
\label{sec:RRback}

In the previous section we have exhibited the BRST structure of the moduli
action for the D(--1)/D7 system of Type I$^\prime$ and found a BRST charge $Q$
that is nilpotent on quantities invariant under the D-instanton group
$\mathrm{SO}(k)$ and the D7 group $\mathrm{SO}(8)$, as shown in (\ref{Q2tt}).
However, since the moduli action is also invariant under the auxiliary group
$\mathrm{SO}(7)$, it is natural to consider an $\mathrm{SO}(7)$-equivariant
cohomology \cite{Moore:1998et}, using a deformed BRST charge that squares to
zero up to an infinitesimal $\mathrm{SO}(7)$ transformation as well. Such a
deformation is the analogue of the $\epsilon$-deformation introduced in
Ref.~\cite{Nekrasov:2002qd} to derive the non-perturbative contributions to the
prepotential of $\mathcal N=2$ super Yang-Mills theories in four dimensions
using localization techniques for the integral over the instanton moduli space
(see for example
Refs.~\cite{Flume:2002az}-\nocite{Bruzzo:2002xf,Nekrasov:2003rj,Bruzzo:2003rw}
\cite{Marino:2004cn}). As shown in Ref.~\cite{Billo:2006jm}, the
$\epsilon$-deformation has a natural interpretation in the string realization of
the instanton calculus since it can be obtained from the interactions of the
open strings of the D(--1)/D3 system with the 3-form field strength of the
Ramond-Ramond (RR) closed string sector representing a constant (self-dual)
graviphoton background. In this section we are going to show that also the
$\mathrm{SO}(7)$ deformation can be obtained by turning on a constant RR
background in the D(--1)/D7 system.

To this aim, let us consider a RR 3-form field strength of type $F_{\mu\nu z}$,
{\it i.e.} with two indices along the 8-dimensional world-volume of the D7
branes and one holomorphic index in the internal torus $T_2$. It is not
difficult to realize that such a field strength survives the orientifold
projection (\ref{orientifoldparity}), since $F_{\mu\nu z}$ is even under the
world-sheet parity $\omega$ (like any other RR 3-form field strength), odd under
$(-1)^{F_L}$ (like any field of the RR sector) and odd under the inversion
$\mathcal I_2$ (like any field with one index in the internal torus). {From} now
on we denote $F_{\mu\nu z}$ simply as $\mathcal F_{\mu\nu}$ and choose it to
describe a rotation of $\mathrm{SO}(7)$ in the spinor representation%
\footnote{In the notation of Appendix \ref{appa} this means that we only turn on
the components $\mathcal F_{\mu\nu}^{\mathbf{21}}$, see \eq{deco28}.} namely we
take
\begin{equation}
 \mathcal F_{\mu\nu} = \frac{1}{2}\,f_{mn} (\tau^{mn})_{\mu\nu}
\label{Ff}
\end{equation}
where $\tau^{mn}=\frac{1}{2}\,\comm{\tau^m}{\tau^n}$ and $f_{mn}$ are the
twenty-one parameters specifying the $\mathrm{SO}(7)$ rotation.

The effects on the moduli action of this RR background can be derived by
computing mixed open/closed string amplitudes on disks with insertions of the
moduli vertex operators on the boundary, and of the vertex operators
representing $\mathcal F$ in the interior. A few details are given in Appendix
\ref{appb} for completeness, but we refer to
Ref.~\cite{Billo:2004zq,Billo:2006jm} for a systematic analysis and a thorough
discussion of this method. In the present case the result of the evaluation of
such mixed amplitudes leads to new couplings in the moduli action which can be
accounted by replacing the cubic and quartic terms, given in (\ref{cubic1}) and
(\ref{quartic1}), as follows
\begin{equation}
\begin{aligned}
 \mathcal S'_{\mathrm{cubic}} & ~\rightarrow~~ 
 \mathcal S'_{\mathrm{cubic}}(\mathcal F) = \mathcal S'_{\mathrm{cubic}} +
 \frac{1}{g_0^2} \,\tr \Big\{\!-\frac 1{2}\,f^{mn}\,\lambda_m\lambda_n\Big\}~,\\
\mathcal S'_{\mathrm{quartic}} & ~\rightarrow~~ 
\mathcal S'_{\mathrm{quartic}}(\mathcal F) = \mathcal S'_{\mathrm{quartic}} +
 \frac{1}{g_0^2} \,\tr \Big\{\frac{\ii}{2\sqrt{2}}\comm{a_\mu}{\bar\chi}
\mathcal F^{\mu\nu}a_\nu 
\Big\} ~.
\end{aligned}
\label{squartic2}
\end{equation}
Thus, when the RR background (\ref{Ff}) is turned on, the moduli action becomes 
\begin{equation}
 \mathcal S'(\mathcal F,\phi)= \mathcal S'_{\mathrm{cubic}}(\mathcal F) +
\mathcal S'_{\mathrm{quartic}}(\mathcal F) + \mathcal S'_{\mathrm{mixed}}(\phi)
\label{smod3}
\end{equation}
with the last term given in (\ref{mixed3}). This new action is still BRST exact,
but with respect to a modified BRST charge $Q^\prime$. Indeed, taking
\begin{equation}
 \label{Q1}
\begin{aligned}
& Q^\prime a^\mu  = M^\mu~,~~~ Q^\prime M^\mu = \ii\sqrt{2}\, 
\comm{\chi}{a^\mu}-\frac12\, \mathcal F^{\mu\nu}\,a_\nu~, \\
& Q^\prime \lambda_m = D_m~,~~~ Q^\prime D_m  =  \ii\sqrt{2}\,
\comm{\chi}{\lambda_m}+f_{mn}\,\lambda^n~,\\
& Q^\prime \bar\chi = -\ii \sqrt{2} \eta~,~~~ 
Q^\prime \eta  = -\comm{\chi}{\bar\chi}~, ~~~ Q^\prime \chi = 0~, \\
& Q^\prime \mu = w~,~~~ Q^\prime w = \ii\sqrt{2}\,\chi\,\mu -
\ii\sqrt{2}\,\mu\,\phi~,
\end{aligned}
\end{equation}
one can check that
\begin{equation}
 \mathcal S'(\mathcal F,\phi) = Q^\prime\,\Xi
\label{SQxi1}
\end{equation}
where the gauge fermion $\Xi$ is the one defined in (\ref{gaugefermion}). The
deformed BRST charge $Q^\prime$ is nilpotent up to (infinitesimal)
transformations of all the symmetry groups of the system, including the
rotations of $\mathrm{SO}(7)$ under which the moduli carrying indices of type
$m,n,...$ (like $\lambda_m$) transform in the vector representation and the
moduli carrying indices of type $\mu,\nu,...$ (like $a^\mu$) transform in the
spinor representation. Indeed, from (\ref{Q1}) one can easily show that
\begin{equation}
 {Q^\prime}^2\,\bullet = T_{\mathrm{SO}(k)}\big(\ii\sqrt{2}\,\chi\big)\bullet
-\,T_{\mathrm{SO}(8)}\big(\ii\sqrt{2}\,\phi\big)\bullet +
\,T_{\mathrm{SO}(7)}\big(\mathcal F\big)\bullet ~.
\label{Q2ttt}
\end{equation}

As discussed in Ref.~\cite{Moore:1998et}, in view of the explicit evaluation of
the integral over the instanton moduli space using localization methods, it is
useful to further deform the above action. Proceeding in strict analogy with
Ref.~\cite{Billo:2006jm}, we turn on also the component of the RR 3-form
field-strength with an anti-holomorphic index, {\it i.e.} $F_{\mu\nu\bar z}
\equiv \bar{\mathcal F}_{\mu\nu}$, and then compute mixed disk amplitudes with
$\bar{\mathcal F}$ insertions to obtain the couplings with the instanton moduli.
Choosing
\begin{equation}
\bar{ \mathcal F}_{\mu\nu} = \frac{1}{2}\,\bar{f}_{mn} (\tau^{mn})_{\mu\nu}
\label{barFf}
\end{equation}
one finds the following new terms
\begin{equation}
 \frac{1}{g_{0}^2}\, \tr
\Big\{\frac{\ii}{2\sqrt{2}}\comm{a_\mu}{\chi}\bar{\mathcal F}^{\mu\nu}a_\nu 
+\frac{1}{8}\,\bar{\mathcal F}^{\mu\nu}a_\nu \,{\mathcal F}_{\mu\rho}a^\rho +
\frac{1}{4} \bar{\mathcal F}_{\mu\nu} M^\mu M^\nu\Big\}
\label{sbarF}
\end{equation}
which have to be added to the moduli action (\ref{smod3}). Notice that the
anti-holomorphic RR background $\bar{\mathcal F}$ produces quadratic ``mass''
terms for the moduli $a^\mu$ and its fermionic partners $M^\mu$.

Another class of deformations which we will use in the following is obtained by
adding to $\mathcal F$ a vector component (see \eq{deco28}), namely by taking
the holomorphic RR polarization tensor to be given by
\begin{equation}
 \mathcal{F}_{\mu\nu}= \frac{1}{2}f_{mn}(\tau^{mn})_{\mu\nu}+ 
h_m(\tau^m)_{\mu\nu}~.
\label{Ffh}
\end{equation}
In this way one gets the following new couplings in the moduli action
\begin{equation}
 \frac{1}{g_{0}^2}\, \tr
\Big\{ h^m \lambda_m \,\eta
+\frac{\ii}{\sqrt 2} \,h^m D_m \bar\chi \Big\}~.
\label{hterms}
\end{equation}
It is important to observe that both the $\bar{\mathcal F}$ terms (\ref{sbarF})
and the $h$ terms (\ref{hterms}) can be incorporated in the BRST structure of
the moduli action by deforming the gauge fermion $\Xi$ and replacing it
according to
\begin{equation}
 \Xi ~\rightarrow~~ \Xi^\prime = \Xi - \frac{1}{g_{0}^2}\, \tr
\Big\{\frac{\ii}{\sqrt{2}}\,h^m \lambda_m\bar\chi
+\frac{1}{4}\,\bar{\mathcal{F}}^{\mu\nu} a_\nu M_\mu \Big\}~.
\label{gaugefermion1}
\end{equation}
Then, the full instanton moduli action in the presence of a RR background given
by (\ref{Ffh}) and (\ref{barFf}) and of a vacuum expectation value $\phi$ for
the adjoint scalar of the gauge multiplet, is given by
\begin{equation}
 \mathcal S'(\mathcal F, \bar{\mathcal F}, \phi) = Q^\prime\,\Xi^\prime~.
\label{smodfin}
\end{equation}
We will take advantage of the BRST exactness of the moduli action in the
following section when we will discuss the integral over the instanton moduli
space.

\section{Rescalings and localization}
\label{sec:scaling}
Our next goal is to compute the instanton partition function for the D(--1)/D7
system using the deformed moduli action derived in the previous section, in
order to extract from it the non-perturbative contributions to the effective
action of the $\mathrm{SO}(8)$ gauge theory. To do so, it is convenient to first
introduce ADHM-like variables by means of the following replacements
\begin{equation}
 a^\mu\,\to\,{a'}^\mu=\frac{a^\mu}{g_0}~,~~~{M}^\mu\,\to
\,{M'}^\mu=\frac{M^\mu}{g_0}~,~~~
\mu\,\to\,\mu'=\frac{\mu}{g_0}~,~~~w\,\to\,w'=\frac{w}{g_0}~,
\label{rescaling}
\end{equation}
in such a way that ${a'}^\mu$ has dimension of (length), ${M'}^\mu$ and $\mu'$
have dimensions of (length)$^{1/2}$, and $w'$ is dimensionless. Then we define
the partition function at instanton number $k$ as the following integral%
\footnote{Here, for simplicity, we do not include the exponential of (minus)
the classical instanton action, $\ee^{2\pi\ii\tau k}$; we will
restore these factors later on.}:
\begin{equation}
 Z_k = \mathcal N_k
\int \{d{a'}^\mu\,d{M'}^\mu\,dD_m\,d\lambda_m\,d\bar\chi\,d\eta\,d\chi\,
d\mu'\,d{w'}\}~
\ee^{-\mathcal S'(\mathcal F,\bar{\mathcal F},\phi)}
\label{Zk}
\end{equation}
where $\mathcal N_k$ is a suitable (dimensionless) normalization factor, and 
$\mathcal S'(\mathcal F,\bar{\mathcal F},\phi)$ is the moduli action
obtained from \eq{smodfin} upon using the rescalings (\ref{rescaling}).

The charged moduli $w'$ and $\mu'$ appear only quadratically in the action
$S'(\mathcal F,\bar{\mathcal F},\phi)$ (see \eq{mixed3}) and can be easily
integrated, yielding%
\footnote{Notice that on the $\mu'$'s the action $Q$ and $Q^\prime$ coincide.}
\begin{equation}
\label{intmu'}
\int \{d\mu'\,d{w'}\} ~\ee^{-\tr \left({}^{\mathrm{t}}w\,w \,+~
{}^{\mathrm{t}}{\!\mu'}{Q^\prime}^2\mu'\right)}
\sim {\mathrm{Pf}}_{(\mathbf{k},\mathbf{8}_v,\mathbf{1})}
\big({Q^\prime}^2\big)
\end{equation}
where the labels on the Pfaffian specify the representations on which
${Q^\prime}^2$ acts. For $k=1$ no $\chi$'s are present and the integral over
$w'$ and $\mu'$ produces just
$\mathrm{Pf}_{(\mathbf{1},\mathbf{8}_v,\mathbf{1})}\big({Q^\prime}
^2\big)\sim\Pf\big(\phi\big)$.

Absorbing all numerical factors into the overall normalization,
we can rewrite the partition function
(\ref{Zk}) as 
\begin{equation}
 Z_k = \mathcal N_k
\int \{d{a'}^\mu\,d{M'}^\mu\,dD_m\,d\lambda_m\,d\bar\chi\,d\eta\,d\chi\}~
\ee^{-\mathcal S'(\mathcal F,\bar{\mathcal F})}
~{\mathrm{Pf}}_{(\mathbf{k},\mathbf{8}_v,\mathbf{1})}
\big({Q^\prime}^2\big)
\label{Zk1}
\end{equation}
where
\begin{equation}
 \begin{aligned}
  \mathcal S'(\mathcal F,\bar{\mathcal F}) =  \tr
\Big\{&\lambda_m (\tau^m)_{\mu\nu}\comm{{a'}^\mu}{{M}'^\nu}+  
\eta \comm{{a'}_\mu}{{M'}^\mu} - \frac{\ii}{\sqrt{2}}\, 
M'_{\mu} \comm{\bar\chi}{{M'}^\mu} \\
& + \frac{1}{2}D_m (\tau^m)_{\mu\nu} \comm{{a'}^{\mu}}{{a'}^{\nu}}
- \comm{a'_\mu}{\bar\chi }\comm{{a'}^\mu}{\chi} \\
&+\frac{\ii}{2\sqrt{2}}\comm{a'_\mu}{\bar\chi}\mathcal F^{\mu\nu}a'_\nu 
-\frac{\ii}{\sqrt{2}g_0^2}\, \eta\comm{\chi}{\eta} + \frac{1}{2g_0^2}\,
\comm{\bar\chi}{\chi}^2 \\
& +\frac{1}{2g_0^2}\,D_m D^m -\frac{1}{2g_0^2}\,\lambda_m\Big(\ii\sqrt{2}
\comm{\chi}{\lambda^m} +f^{mn}\lambda_n\Big)\\
&+\frac{1}{4}\,a'_\mu\,\bar{\mathcal F}^{\mu\nu}
\Big(\ii\sqrt{2}\comm{\chi}{a'_\nu}
-\frac{1}{2}{\mathcal F}_{\nu\rho}\,{a'}^\rho\Big)+\frac{1}{4}
\bar{\mathcal F}_{\mu\nu} {M'}^\mu {M'}^\nu\\
&+ \frac{1}{g_0^2}\,h^m 
\big(\lambda_m\,\eta+\frac{\ii}{\sqrt{2}}\,D_m\bar\chi\big)\Big\}~.
 \end{aligned}
\label{sffbar}
\end{equation}
As customary in this type of manipulations \cite{Nekrasov:2002qd}, we treat the
variables $\chi$ and $\bar\chi$ as independent of each other and, in particular,
according to our conventions, we take them to be purely imaginary and real
respectively. Then, we evaluate the integral (\ref{Zk1}) in the semi-classical
approximation, which due to the BRST structure of the instanton action turns out
to be exact. To proceed it is convenient to perform the following change of
integration variables
\begin{equation}
\begin{aligned}
 &{a'}^\mu\,\to\,\frac{1}{x}\,{a'}^\mu~&,&~~~~{M'}^\mu\,\to\,
\frac{1}{x}\,{M'}^\mu~,\\
 &D_m\,\to\,{x}^2\,D_m~&,&~~~~\lambda_m\,\to\,{x}^2\,\lambda_m~,\\
&\bar\chi\,\to\,{y}\,\bar\chi~&,&~~~~\eta\,\to\,{y}\,\eta~,
\end{aligned}
\label{rescalings}
\end{equation}
and rescale the anti-holomorphic background as
\begin{equation}
 \bar{\mathcal F}_{\mu\nu}\,\to\,z\,\bar{\mathcal F}_{\mu\nu}~.
\label{tFbar}
\end{equation}
The partition function $Z_k$ does not depend on the arbitrary parameters $x$,
$y$ and $z$, because $x$ and $y$ appear only through a change of integration
variables which leaves invariant the measure in (\ref{Zk1}), while $z$ appears
through a change of the anti-holomorphic background which only appears inside
the gauge fermion $\Xi^\prime$ as shown in (\ref{gaugefermion1}). Thus, we can
choose these parameters to simplify as much as possible the structure of $Z_k$.
In particular, if we take the limit
\begin{equation}
 x\,\to\,\infty\quad,\quad y\,\to\,0\quad,\quad z\,\to\,\infty
\label{limits}
\end{equation}
with
\begin{equation}
 x^2y\,\to\,\infty\quad,\quad \frac{z}{x^2}\,\to\,\infty~,
\label{conditions}
\end{equation}
the moduli action (\ref{sffbar}) reduces to
\begin{equation}
 \label{simplaction}
\begin{aligned}
\mathcal S'(\mathcal F,\bar{\mathcal F}) =\,& \tr \Big\{\frac{g}{2}\,D_m D^m -
\frac{g}{2}\,\lambda_m\big(\ii\,\sqrt{2}\comm{\chi}{\lambda^m}+f^{mn}\,
\lambda_n\big)\\
&+\frac{t}{4}\,a'_\mu\,\bar{\mathcal F}^{\mu\nu}
\big(\ii\,\sqrt{2}\comm{\chi}{a'_\nu}-\frac{1}{2}
{\mathcal F}_{\nu\rho}\,{a'}^\rho\big)
+\frac{t}{4}\,M'_\mu\,\bar{\mathcal F}^{\mu\nu}M'_\nu\\
&+s\,h^m\big(\lambda_m\,\eta+\frac{\ii}{\sqrt{2}}\,D_m\bar\chi\big)
\Big\} + \ldots~.
\end{aligned}
\end{equation}
Here we have introduced the coupling constants
\begin{equation}
 g=\frac{x^4}{g_0^2}~,~~~~ t= \frac{z}{x^2}~,~~~~ s=\frac{x^2y}{g_0^2}~,
\label{couplings}
\end{equation}
which all tend to $\infty$ because of \eq{conditions}, and have denoted with
$\ldots$ the terms of the first three lines of \eq{sffbar} which are subleading
in this limit. The integrals over ${a'}^\mu$, ${M'}^\mu$, $D_m$, $\lambda_m$,
$\bar\chi$ and $\eta$ can now be easily performed since they are all Gaussian. 

To evaluate these integrals we choose the deformation parameters $f_{mn}$ and
$h_m$ as in Ref.~\cite{Moore:1998et}, namely to take the matrix $f$ along the
Cartan directions $H^a_{\mathrm{SO}(7)}$ of $\mathrm{SO}(7)$, {\it i.e.}
\begin{equation}
\label{fmatrix}
f = \vec{f}\cdot \vec{H}_{\mathrm{SO}(7)} = 
\sum_{a=1}^3\, f_a\,H^a_{\mathrm{SO}(7)} 
= \begin{pmatrix} 
      \ii f_1\sigma_2& 0 &0&0  \cr
     0&\ii f_2\sigma_2&0&0 \cr
      0&0&\ii f_3\sigma_2& 0\cr
      0&0&0&0 
     \end{pmatrix}
\quad\mbox{with}~\sigma_2= \begin{pmatrix} 
     0& -\ii  \cr
     \ii &0     \end{pmatrix} ~,
\end{equation}
and the vector $h$ with only $h_7$ non-vanishing%
\footnote{Even if this is not the most general configuration, it is the most
convenient one for the following computations.}. When these parameters are
inserted in (\ref{simplaction}), the fermion $\lambda_7$ lacks an explicit
``mass term'' from the $(f\lambda\lambda)$ coupling but it becomes effectively
``massive'' thanks to the $(\lambda_7\eta)$ term proportional to $h_7$ and thus
can be integrated without problems. Actually, it is easy to integrate out the
entire quartet formed by $D_7$, $\lambda_7$, $\bar\chi$ and $\eta$ and realize
that it yields just a numerical constant independent of $g$, $s$, $h_7$ and
$\chi$. Indeed, even if these quantities do appear in the interactions among the
quartet components, they can be scaled away by a change of integration variables
that leaves the integration measure invariant.

Once the quartet has been integrated, we can safely set $h_7=0$. Thus, the 
deformation matrix (\ref{Ffh}) becomes
\begin{equation}
\label{Ff1}
\mathcal F =-2 \begin{pmatrix} 
      \ii E_1\sigma_2& 0 &0&0  \cr
     0&\ii E_2\sigma_2&0&0 \cr
      0&0&\ii E_3\sigma_2& 0\cr
      0&0&0&\ii E_4\sigma_2 
     \end{pmatrix} 
\end{equation}
with
\begin{equation}
 \begin{aligned}
E_1&=\frac{1}{2}\big(f_1-f_2-f_3\big)\quad,\quad
E_2=\frac{1}{2}\big(f_2-f_3-f_1\big)~,\\
E_3&=\frac{1}{2}\big(f_3-f_1-f_2\big)\quad,\quad
E_4=\frac{1}{2}\big(f_1+f_2+f_3\big)~,
 \end{aligned}
\label{fE}
\end{equation}
such that
\begin{equation}
 E_1+E_2+E_3+E_4=0~.
\label{sumE}
\end{equation}

At this point, we are left with the integral over ${a'}^\mu$, ${M'}^\mu$, the
six ``massive'' fermions $\lambda_1,\ldots,\lambda_6$ (which we will label with
an index $\hat m=1,\ldots,6$) and the corresponding six auxiliary bosons
$D_{\hat m}$, plus of course the integral over $\chi$. {From} \eq{simplaction},
we see that the relevant action for these fields is extremely simple and given
by
\begin{equation}
 \label{simplaction1}
\tr \Big\{\frac{g}{2}\,D_{\hat m} D^{\hat m} -
\frac{g}{2}\,\lambda_{\hat m}\, \big({Q'}^2\lambda\big)^{\hat m}
+\frac{t}{4}\,a'_\mu\,\bar{\mathcal F}^{\mu\nu}\big({Q'}^2a'\big)_\nu
+\frac{t}{4}\,M'_\mu\,\bar{\mathcal F}^{\mu\nu}M'_\nu\Big\}
\end{equation}
with the deformed BRST charge acting as in \eq{Q2ttt}. The integral we have to
compute is then
\begin{equation}
\label{int2}
\begin{aligned}
 I =\int \big\{d{a'}^\mu d{M'}^\mu & dD_{\hat m}d\lambda_{\hat m}\big\}\,
\ee^{-\tr \big\{\frac{g}{2}D_{\hat m} D^{\hat m} -
\frac{g}{2}\lambda_{\hat m} ({Q'}^2\lambda)^{\hat m}
+\frac{t}{4}a'_\mu\bar{\mathcal F}^{\mu\nu}({Q'}^2a')_\nu
+\frac{t}{4}M'_\mu\bar{\mathcal F}^{\mu\nu}M'_\nu\big\}}\\
&\sim ~ \frac{\Pf_{(\mathrm{adj},\mathbf{1},
{\mathbf 6}\subset{\mathbf 7})}{(g\,{Q'}^2)}
{\phantom{\Big|}}
~\Pf_{(\mathrm{symm},\mathbf{1},\mathbf{8}_s)}(\frac{t}{2}\,\bar{\mathcal F})}
{\det^{1/2}_{\,(\mathrm{adj},{\mathbf{1},\mathbf 6}\subset{\mathbf 7})}(g) 
{\phantom\vdots}\det^{1/2}_{\,(\mathrm{symm},\mathbf{1},\mathbf{8}_s)}
(\frac{t}{2}\,\bar{\mathcal F}{Q'}^2)}
\end{aligned}
\end{equation}
up to numerical coefficients. The origin of the various terms in the above
expression is clear: $\Pf_{(\mathrm{adj},\mathbf{1},{\mathbf 6}\subset{\mathbf
7})}{(g\,{Q'}^2)}$ comes from the integration of the six fermions $\lambda_{\hat
m}$ which transform in the adjoint representation of $\mathrm{SO}(k)$, are
singlets of $\mathrm{SO}(8)$ and form a 6-vector inside the ${\mathbf 7}$ of
$\mathrm{SO}(7)$, as indicated by the labels on the Pfaffian symbol. Similarly,
$\Pf_{(\mathrm{symm},\mathbf{1},\mathbf{8}_s)}(\frac{t}{2}\,\bar{\mathcal F})$
comes from the integration of the fermions ${M'}^\mu$;
$\det^{1/2}_{\,(\mathrm{adj},\mathbf{1},{\mathbf 6}\subset{\mathbf 7})}(g)$
comes from the integration of the six bosons $D_{\hat m}$ and finally
$\det^{1/2}_{\,(\mathrm{symm},\mathbf{1},\mathbf{8}_s)}(\frac{t}{2}\,\bar{
\mathcal F}{Q'}^2)$ comes from the integration of the bosons ${a'}^\mu$.
Exploiting the properties of the Pfaffians, we can simplify \eq{int2} and get
\begin{equation}
 \label{int3}
I \sim~\frac{\Pf_{(\mathrm{adj},\mathbf{1},{\mathbf 6}\subset{\mathbf
7})}{({Q'}^2)} {\phantom{\Big|}}}
{{\phantom\vdots}\det^{1/2}_{\,(\mathrm{symm},\mathbf{1},\mathbf{8}_s)}({Q'}^2)}
~.
\end{equation}
As expected, all dependence on $g$, $t$ and the anti-holomorphic background
$\bar{\mathcal F}$ has dropped out from the final result, which instead depends
on the holomorphic background $\mathcal F$ given in (\ref{Ff1}) and on $\chi$
(the last instanton moduli to be integrated) through the action of the deformed
BRST charge.

Combining everything and absorbing all numerical factors in the overall
normalization coefficient, we finally obtain
\begin{equation}
 \label{Zk3}
 Z_k = \mathcal N_k
\int \{d\chi\}~\frac{\Pf_{(\mathrm{adj},\mathbf{1},{\mathbf 6}\subset{\mathbf
7})}{({Q'}^2)} {\phantom{\Big|}}\Pf_{(\mathbf{k},\mathbf{8}_v,{\mathbf
1})}{({Q'}^2)}}{{\phantom\vdots}\det^{1/2}_{\,(\mathrm{symm},
\mathbf{1},\mathbf{8}_s)}({Q'}^2)} ~.
\end{equation}
As suggested by \eq{Q2ttt}, it is convenient to redefine
$\ii\sqrt{2}\chi\,\to\,\chi$ and $\ii\sqrt{2}\phi\,\to\,\phi$, so that
the new $\chi$ variable becomes real and
\begin{equation}
 {Q^\prime}^2\,\bullet = T_{\mathrm{SO}(k)}\big(\chi\big)\bullet
-\,T_{\mathrm{SO}(8)}\big(\phi\big)\bullet +
\,T_{\mathrm{SO}(7)}\big(\mathcal F\big)\bullet ~.
\label{Q2ttt1}
\end{equation}
Furthermore, for ease of notation we set
\begin{equation}
 \label{notationpf}
\cP(\chi) \equiv \Pf_{(\mathrm{adj},\mathbf{1},
{\mathbf 6}\subset{\mathbf 7})}({Q'}^2)~,~ \cR(\chi) \equiv
\Pf_{(\mathbf{k},\mathbf{8}_v,{\mathbf 1})}({Q'}^2)~,~
\cQ(\chi) \equiv
\det{}^{1/2}_{\,(\mathrm{symm},\mathbf{1},\mathbf{8}_s)}({Q'}^2)
\end{equation}
and, after a suitable redefinition of the overall normalization, we rewrite the
partition function as follows
\begin{equation}
Z_k = \cN_k
\int \Big\{\frac{d\chi}{2\pi\ii}\Big\}~\frac{\cP(\chi)\, \cR(\chi)}{\cQ(\chi)}~.
\label{Zk2}
\end{equation}
Since the integrand is singular when the denominator $\cQ(\chi)$ vanishes and
tends to one when $\chi\to\infty$, the integral (\ref{Zk2}) is naively divergent
and must be suitably defined to make sense. Here we follow the same prescription
of Ref.~\cite{Moore:1998et}, and cure the singularities along the integration
path by giving the zeroes of $\cQ(\chi)$ a small positive imaginary part moving
them in the upper-half complex plane, and regulate the divergence at infinity by
interpreting the $\chi$-integral as a contour integral. Even if this
prescription as it stands does not seem to be fully justified and lacks a
rigorous derivation from first principles, there is clear evidence of its
validity in results of Ref.~\cite{Moore:1998et} and their numerous
generalizations discussed for example in
Refs.~\cite{Krauth:1998yu}-\nocite{Krauth:2000bv,Staudacher:2000gx,
Pestun:2002rr}\cite{Fischbacher:2003un}, as well as in the agreement with
numerical analysis based on Monte-Carlo methods \cite{Krauth:1998xh}.

Using this prescription, the instanton partition function (\ref{Zk2}) will then
be expressed as a finite sum of residues evaluated at the poles of the
integrand, showing that the integral over the instanton moduli effectively
localizes on the zeroes of $\cQ(\chi)$ and thus receives contributions only from
those configurations for which the bosonic ``kinetic'' terms vanish. This is
completely similar to the localization of the integrals over the instanton
moduli space in $\mathcal N=2$ super Yang-Mills theories in four dimensions
discussed in Ref.s~\cite{Nekrasov:2002qd} and
\cite{Flume:2002az}-\nocite{Bruzzo:2002xf,Nekrasov:2003rj,Bruzzo:2003rw}\cite{
Marino:2004cn}.

\section{Explicit expressions and results for low $k$}
\label{sec:explicit}

\subsection{$k=1$}
The 1-instanton partition function $Z_1$ is particularly simple: in fact, for
$k=1$ there are no $\lambda_m$'s and no $\chi$'s, so that the factor $\cP(\chi)$
is not generated and no contour integral has to be evaluated. Furthermore, for
$k=1$ the factor $\cR(\chi)$ reduces just to $\Pf\big(\phi\big)$, as already
observed after \eq{intmu'}, while from \eq{int2} we see that the integration
over ${a'}^\mu$ and ${M'}^\mu$ reduces to
\begin{equation}
 \int \big\{d{a'}^\mu d{M'}^\mu\big\}\,\ee^{-\big\{
\frac{1}{8}a'_\mu\bar{\mathcal F}^{\mu\nu}\cF_{\rho\nu}{a'}^\rho
+\frac{1}{4}M'_\mu\bar{\mathcal F}^{\mu\nu}M'_\nu\big\}} ~\sim~
\frac{1}{\det^{1/2}(\cF)} ~\sim~\frac{1}{\cE} 
\label{intam1}
\end{equation}
where have defined
\begin{equation}
\label{PfE}
 \cE\,\equiv \, E_1 E_2 E_3 E_4~.
\end{equation}
Thus, for $k=1$ we simply have
\begin{equation}
 Z_1 = \cN_1\,\frac{\Pf\phi}{\cE}~.
\label{Z1fin}
\end{equation}
Notice that the factor $1/\cE$ in the above result can be interpreted as the
regulated volume of the eight-dimensional $\cN=1$ superspace. In fact, for $k=1$
the moduli ${a'}$ and ${M'}$ are identified with the superspace coordinates (see
\eq{xtheta}), so that from (\ref{intam1}) we can obtain the effective
identification%
\footnote{The factors of $\pi$'s are introduced for later convenience, but it is
easy to trace their origin in the Gaussian integration over the eight bosonic
moduli ${a'}^\mu$.}
\begin{equation}
 \int d^8x\,d^8\theta ~\longleftrightarrow~\frac{(2\pi)^4}{\cE}~.
\label{supervol} 
\end{equation}
This is the eight-dimensional analogue of the effective rule that appears in the
instanton calculus in four dimensions using localization and
$\epsilon$-deformation methods \cite{Nekrasov:2002qd,Billo:2006jm}.

\subsection{$k>1$}
Let us now consider the cases with $k>1$. To perform the integration over the
$\chi$'s we can exploit the $\mathrm{SO}(k)$ invariance of the integrand in
(\ref{Zk2}) and, at the price of introducing a Vandermonde determinant%
\footnote{Notice that this operation is formally acceptable only when $\chi$ is
real, which is what we have argued at the end of the previous section.}
$\Delta(\chi)$, bring the $\chi$'s to the Cartan subalgebra, 
whose generators we denote as $H^i_{\mathrm{SO}(k)}$, {\it i.e.}
\begin{equation}
 \label{chicartan}
 \chi ~~\to~~ \vec\chi \cdot \vec H_{\mathrm{SO}(k)}~= 
\sum_{i=1}^{\mathrm{rank}\, \mathrm{SO}(k)}
\chi_{i} \,H^{i}_{\mathrm{SO}(k)} ~.
\end{equation}
Then the partition function becomes
\begin{equation}
\label{Zkred}
Z_k = \cN_k \int \prod_{i} \Big(\frac{d \chi_{i}}{2\pi\ii}\Big)~ 
\Delta(\vec\chi)\, \frac{\cP(\vec \chi)\, \cR(\vec\chi)}{\cQ(\vec\chi)}~.
\end{equation}
Again, we have absorbed all numerical factors produced by the
``diagonalization'' of $\chi$ into a redefinition of the normalization
coefficient $\cN_k$.

Without any loss of generality we can assume that also the vacuum expectation
values of the scalar $\phi$ belong to the Cartan directions
$H^u_{\mathrm{SO}(8)}$
of $\mathrm{SO}(8)$ and thus have the following block-diagonal form 
\begin{equation}
\label{cartanphi}
\phi = \vec{\phi}\cdot \vec{H}_{\mathrm{SO}(8)} = \sum_{u=1}^4\, 
\phi_u\,H^u_{\mathrm{SO}(8)} 
= \begin{pmatrix} 
      \ii \phi_1\sigma_2& 0 &0&0  \cr
     0&\ii \phi_2\sigma_2&0&0 \cr
      0&0&\ii \phi_3\sigma_2& 0\cr
      0&0&0&\ii \phi_4\sigma_2 
     \end{pmatrix} ~.
\end{equation}
With these choices, ${Q^\prime}^2$ corresponds to infinitesimal Cartan actions
which can be diagonalized in any representation by going to the basis provided
by the weights. 

Let consider, for instance, the charged moduli $\mu'$ which we relabel as 
\begin{equation}
 \label{relabl}
{\mu'}^I_{~U} ~\to~ {\mu'}^{\vec\pi}_{~\vec\gamma}~\sim \ket{\vec\pi,\vec\gamma}~,
\end{equation} 
where $\vec\pi$ belongs to the set of weights of the vector representation
$\mathbf{k}$ of $\mathrm{SO}(k)$, while $\vec\gamma$ is a weight of the vector
representation $\mathbf{8}_v$ of $\mathrm{SO}(8)$. Then, from (\ref{Q2ttt1}) we
have
\begin{equation}
 \label{chicartis}
\begin{aligned}
{Q^\prime}^2\,\ket{\vec\pi,\vec\gamma} = 
\left(T_{\mathrm{SO}(k)}(\vec\chi) - T_{\mathrm{SO}(8)}(\vec\phi)\right)
\ket{\vec\pi,\vec\gamma} & = \big(\vec\chi\cdot \vec\pi - \vec\phi\cdot
\vec\gamma\big)\ket{\vec\pi,\vec\gamma}~.
\end{aligned}
\end{equation}
Notice that the variables ${\mu'}^{\vec\pi}_{~\vec\gamma}$ are in general
complex, and their conjugate moduli are%
\footnote{All representations appearing in our expressions are real, namely
correspond to weight sets that are closed under parity.}
${\mu'}^{-\vec\pi}_{~-\vec\gamma}$; the couples of conjugate moduli are
therefore labeled by half of the possible pairs of weights
$(\vec\pi,\vec\gamma)$. Hence, the complex fermionic integration over the
${\mu'}^{\vec\pi}_{~\vec\gamma}$'s yields 
\begin{equation}
 \label{pfchiPhiwe}
\cR(\vec\chi) = 
\prod_{\vec\pi\in \mathbf{k}} \prod^{(+)}_{\vec\gamma\in \mathbf{8}_v} \left(
\vec\chi\cdot\vec\pi - \vec\phi\cdot\vec\gamma\right)~.
\end{equation}
Here the product over $\vec\gamma$ is limited to half of the weights, that we
refer to as the ``positive'' ones; this is the meaning of the superscript $(+)$
appearing above. The weights of the vector representation $\mathbf{8}_v$ of
$\mathrm{SO}(8)$ are expressed in terms of the versors $\ve u$ ($u=1,\ldots 4$)
spanning the weight space as $\pm\ve u$. Taking $\ve u$ as the positive ones, we
obtain 
\begin{equation}
 \label{pfchiPhi1}
\cR(\vec\chi) =
\prod_{u=1}^4 \prod_{\vec\pi\in \mathbf{k}} 
\Big(\vec\chi\cdot\vec\pi - \phi_u\Big)~.
\end{equation}
We can proceed in a similar way for the six moduli $\lambda_{\hat m}$, finding 
\begin{equation}
 \label{pflambdaF}
\cP(\vec\chi) =
 \prod_{\vec\rho\in \mathrm{adj}} 
\prod^{(+)}_{\vec\alpha\in \mathbf{6}\subset \mathbf{7}} \left(
\vec\chi\cdot\vec\rho - \vec f\cdot\vec\alpha\right)
= \prod_{a=1}^3  \prod_{\vec\rho\in \mathrm{adj}} 
\Big( \vec\chi\cdot\vec\rho - f_a\Big)~.
\end{equation}
Indeed, the positive weights of the $\mathbf{6}\subset \mathbf{7}$
representation of $\mathrm{SO}(7)$ correspond simply to the versors $\ve a$
($a=1,2,3$) of the weight space. Finally, considering the moduli ${a'}^\mu$, we
get
\begin{equation}
 \label{detchiF}
\cQ(\vec\chi)=
\prod_{\vec\sigma\in \mathrm{symm}} \prod^{(+)}_{\vec\beta\in \mathbf{8}_s} 
\left(\vec\chi\cdot\vec\sigma - \vec f\cdot\vec\beta\right)
= \prod_{A=1}^4 \prod_{\vec\sigma\in \mathrm{symm}} \Big(\vec\chi\cdot\vec\sigma
- E_A\Big)~.
\end{equation}
Here $E_A$ ($A=1,\ldots 4$) denote the scalar products of the background $\vec
f$ given in (\ref{fmatrix}) with the four positive weights of the spinor
representation of $\mathrm{SO}(7)$, and correspond precisely to the parameters
introduced in (\ref{fE}). Also the Vandermonde determinant can be expressed in
terms of the non-zero weights of the adjoint representation of SO$(k)$:
\begin{equation}
 \label{vexp}
\Delta(\vec\chi) = \prod_{\vec\rho\in \mathrm{adj}\not=\vec 0} 
\vec\chi\cdot\vec\rho~.
\end{equation}

All the above expressions become explicit using the weight sets of the various
representations provided in Appendix \ref{app:detsok}. As an illustration, let
us discuss $\cR(\vec\chi)$ given in (\ref{pfchiPhi1}). When $k=2n$, the rank of
$\mathrm{SO}(k)$ is $n$. Denoting the versors of the $\mathbb{R}^n$ weight space
as $\ve{i}$, the weights $\vec\pi$ of the vector representation $\mathbf{2n}$
are simply $\vec\pi = \pm\ve i$, so that from (\ref{pfchiPhi1}) we get
\begin{equation}
 \label{Rnull}
\cR(\vec\chi) =
\prod_{u=1}^4\prod_{i=1}^n  (\chi_i - \phi_u)(-\chi_i - \phi_u)
=\prod_{u=1}^4 \prod_{i=1}^n (\phi_u^2 -\chi_i^2)~.
\end{equation}
For $k=2n+1$, the rank of $\mathrm{SO}(k)$ is again $n$ but now the vector
representation contains in addition to the weights $\pm\ve i$ also a null weight
$\vec 0$. As a consequence, we find an extra factor of $\Pf\phi$; indeed
\begin{equation}
 \label{Rnullodd}
\cR(\vec\chi) =
\prod_{u=1}^4 (-\phi_u) \prod_{i=1}^n (\chi_i - \phi_u)(-\chi_i - \phi_u)
=\Pf\phi\,\prod_{u=1}^4 \prod_{i=1}^n (\phi_u^2 -\chi_i^2)~.
\end{equation}
Let us notice that also the adjoint and symmetric representations of SO$(k)$
contain null weights, which lead to terms independent of $\vec\chi$ in the
products (\ref{pflambdaF}) and (\ref{detchiF}). In particular, the adjoint
representation has $n$ null weights both for $k=2n$ and $k=2n+1$, leading to 
\begin{equation}
 \label{Pnull}
\cP(\vec\chi)
 = \big(\Pf f\big)^n   \prod_{a=1}^3 \prod_{\vec\rho\in \mathrm{adj}\not=\vec 0} 
 \left( \vec\chi\cdot\vec\rho - f_a\right)~,
\end{equation}
where 
\begin{equation}
\label{Pff}
 \Pf f = f_1 f_2 f_3~.
\end{equation}
The symmetric representation, instead, 
has $n$ null weights when $k=2n$, and $n+1$ when $k=2n+1$, so that 
\begin{equation}
 \label{Qnulleven}
\cQ(\vec\chi) =
 \mathcal{E}^{n} \,\prod_{A=1}^4 \prod_{\vec\sigma\in \mathrm{symm}\not=\vec 0} 
 \left(\vec\chi\cdot\vec\sigma - E_A\right)~~
 \mbox{for $k=2n$}
\end{equation}
and
\begin{equation}
 \label{Qnullodd}
 \cQ(\vec\chi) =
 \mathcal{E}^{n+1} \,\prod_{A=1}^4 \prod_{\vec\sigma\in \mathrm{symm}\not=\vec 0} 
 \left(\vec\chi\cdot\vec\sigma - E_A\right)~~
 \mbox{for $k=2n+1$}~,
\end{equation}
where $\cE$ is the quantity defined in (\ref{PfE}).

Using these explicit expressions we can perform the final integrations over the
$\chi$'s and obtain the instanton partition functions $Z_k$ given in
(\ref{Zkred}). As discussed at the end of Section \ref{sec:scaling}, the
$\chi$-integrals are understood as contour integrals in the upper-half complex
plane and the singularities at the zeroes of the polynomial $\cQ(\vec\chi)$ are
avoided by giving the deformation parameters $E_A$ a small positive imaginary
part, according to the prescriptions of Ref.~\cite{Moore:1998et}. In particular,
we choose
\begin{equation}
 \label{imparts}
 \im E_1 > \im E_2 > \im E_3 > \im E_4 > \im \frac{E_1}{2} > \ldots > 
\im \frac{E_4}{2} > 0~.
\end{equation}

Let us apply this to the simplest non-trivial case, namely $k=2$, where we have
\begin{equation}
\label{Z2}
Z_2 = \cN_2 \,\frac{\Pf f}{\mathcal{E}}
\int \frac{d\chi}{2\pi\ii} ~
\frac{\prod_{u=1}^4 (\phi_u^2 - \chi^2)}{\prod_{A=1}^4 
(2\chi - E_A)(-2\chi -E_A)}~.
\end{equation}
The integration prescription described above leads to express $Z_2$ as a sum
over the residues of the integrand at $\chi = E_A/2$: 
\begin{equation}
 \label{Z2res}
Z_2 = \cN_2 \,\frac{\Pf f}{2\,\mathcal{E}} \,\sum_{A=1}^4 
\frac{\prod_{u=1}^4 (\phi_u^2 - (E_A/2)^2)}{\prod_{B\not= A}
(E_A - E_B) \prod_B (-E_A - E_B)}~.
\end{equation}
If we perform the algebra, and use the relations (\ref{fE}) between the
quantities $E_A$ and the three independent parameters $f_a$, in the end we get 
\begin{equation}
 \label{Z2ris}
\begin{aligned}
 Z_2 = \cN_2 \,
\Bigg\{\frac{\big(\Pf\phi\big)^2}{4\, \mathcal{E}^2} 
+ \frac{1}{\mathcal{E}} 
\Bigg[&\frac{\Tr\phi^4 - \frac 12 \big(\Tr\phi^2\big)^2}{256}    \\
& +\frac{\Tr f^2\, \Tr \phi^2}{2048} 
+\frac{\Tr f^4- \frac 54 \big(\Tr f^2\big)^2}{16384} \Bigg]\Bigg\}~.
\end{aligned}
\end{equation}
Here we have rewritten the resulting polynomials in the eigenvalues $f_a$ and
$\phi_u$ in terms of invariants constructed with the matrices $\phi$ and $f$ in
order to get expressions that, although derived choosing $\phi$ and $f$ in the
Cartan directions, are valid generically. For instance, the terms of order
$\phi^4$ in (\ref{Z2ris}) arise in the form
\begin{equation}
 \label{firstform}
  \sum_{u>v}\phi_u^2\phi_v^2 = -\frac 14\Big(\Tr \phi^4 - 
  \frac 12 \big(\Tr\phi^2\big)^2\Big)~,
\end{equation}
since, according to \eq{cartanphi}, in the block-diagonal case we have
\begin{equation}
 \label{t2traces}
 \Tr\phi^2 = -2 \sum_u \phi_u^2~,~~~ \Tr\phi^4 = 2 \sum_u \phi_u^4~.
\end{equation}

For $k=3$, the integral to be computed reads
\begin{equation}
\label{Z3}
Z_3 = \cN_3  \,\frac{\Pf\phi \,\Pf f}{\mathcal{E}^2}
\int \frac{d\chi}{2\pi\ii} ~\frac{\chi^2 \prod_{u=1}^4 (\phi_u^2 -\chi^2)
\prod_{a=1}^3 (f_a^2 - \chi^2)}{\prod_{A=1}^4
 (2\chi - E_A) (-2\chi - E_A)(\chi - E_A)(-\chi - E_A)}~.
\end{equation} 
The integration prescription leads now to the sum over two classes of residues,
those in $\chi = E_A/2$ and those in $\chi = E_A$. After the algebra has been
carried out, this sum reduces to 
\begin{equation}
\label{solZ3}
\begin{aligned}
Z_3  = \cN_3\, \Pf\phi\,
\Bigg\{ \frac{\big(\Pf\phi\big)^2}{12 \,\mathcal{E}^3} + \frac{1}{\mathcal{E}^2}
\Bigg[&\frac{\Tr\phi^4 - \frac 12 \big(\Tr\phi^2\big)^2}{256} 
+ \frac{\Tr f^2\, \Tr\phi^2}{2048}  \\
&  + \frac{\Tr f^4 - \frac 54 \big(\Tr f^2\big)^2}{16384} 
\Bigg] + \frac{1}{96 \mathcal{E}}
\Bigg\}~.
\end{aligned}
\end{equation}

In the cases $k=4$ and $k=5$ the rank of SO$(k)$ equals 2 and we have therefore
to perform a double contour integral over $\chi_1$ and $\chi_2$. In Appendix
\ref{subapp:k45} we give some details about the classes of residues that
contribute to these integrations. The  complete resulting expressions for $Z_4$
and $Z_5$ are too cumbersome to report them explicitly; however, we report the
terms with the highest power of $\cE$ in the denominator, namely
\begin{subequations}
 \begin{align}
  Z_4 &=\cN_4\,\frac{\big(\Pf\phi\big)^4}{48\,\mathcal{E}^4} + 
  \cdots~,\label{z4a}\\
  Z_5 &= \cN_5\,
\frac{\big(\Pf\phi\big)^5}{240\,\mathcal{E}^5} + \cdots ~,
\label{z5a}
 \end{align}
\end{subequations}
which will be useful for the calculations described in the next section.

\section{The prepotential and its gravitational corrections}
\label{sec:prep}
{From} the instanton partition functions $Z_k$ computed in the previous section,
we define the ``grand-canonical'' partition function
\begin{equation}
 \label{Ztot}
\mathcal Z =
\sum_{k=0}^\infty Z_k\, \ee^{2\pi\ii\tau k} =
\sum_{k=0}^\infty Z_k\, q^k
\end{equation}
where we have conventionally set $Z_0=1$, and, as in (\ref{qdef}), defined $q
\equiv\exp(2\pi\ii\tau)$. This allows us to obtain the non-perturbative
contributions to the effective action of the D7-branes. However, to do so one
has first to take into account the fact that the $k$-th order in the
$q$-expansion receives contributions not only from genuine $k$-instanton
configurations but also from ``disconnected'' ones, corresponding to copies of
instantons of lower numbers $k_i$ such that $\sum k_i = k$
\cite{Nekrasov:2002qd}. Thus, to isolate the connected components we have to
take the logarithm of $\mathcal Z$. Moreover, as we have explicitly shown in the
previous sections, the partition functions $Z_k$ have been obtained by
integrating over {\it all} moduli, including the ``center of mass'' coordinates
$x^\mu$ and their superpartners $\theta^\alpha$ defined in (\ref{xtheta}). In
absence of deformations these zero-modes do not appear in the moduli action and
the integration over them would diverge, producing the (infinite)
``supervolume'' of the eight-dimensional base manifold. In presence of SO$(7)$
deformations, instead, as we remarked around \eq{supervol}, the integration over
the superspace coordinates yields a factor of $(2\pi)^4/\mathcal{E}$. Therefore,
to obtain the integral over the centered moduli only, it is sufficient to remove
this factor. Having done so, we can promote the vacuum expectation value $\phi$
appearing in $\mathcal Z$ to the full fledged dynamical superfield
$\Phi(x,\theta)$ and, after removing the RR deformation, obtain the
non-perturbative contributions to the effective action of the D7-branes, namely
\begin{equation}
 \label{SeffPhi}
 S^{\mathrm{(n.p.)}} = \frac{1}{(2\pi)^4}\int d^8x\, d^8\theta\, 
F^{\mathrm{(n.p.)}}\big(\Phi(x,\theta)\big)
\end{equation}
with the ``prepotential'' $F^{\mathrm{(n.p.)}}(\Phi)$ given by
\begin{equation}
 \label{prep1}
 F^{\mathrm{(n.p.)}}(\Phi) = {\mathcal{E}}\, 
\log \mathcal Z\Big|_{\phi\to\Phi,f\to 0}~.
\end{equation}
Expanding in instanton contributions we can write
\begin{equation}
 \label{Fexp}
 F^{\mathrm{(n.p.)}}(\Phi) = \sum_{k=1}^\infty F_k\, q^k~
\Big|_{\phi\to\Phi,f\to 0}
\end{equation}
and, using (\ref{Ztot}), express recursively each $F_k$ in terms of the
partition functions $Z_k$ and of the coefficients $F_j$ with $j<k$, according
to 
\begin{equation}
 \label{Fkexp}
 \begin{aligned}
  F_1 & = \cE Z_1~,\\
  F_2 & = \cE Z_2 - \frac {F_1^2}{2\cE}~,\\
  F_3 & = \cE Z_3 - \frac{F_2 F_1}{\cE} - \frac{F_1^3}{6\cE^2}~,\\
  F_4 & = \cE Z_4 - \frac{F_3 F_1}{\cE} - \frac{F_2^2}{2\cE} -
          \frac{F_2 F_1^2}{2\cE^2} - \frac{F_1^4}{24\cE^3}~,\\
  F_5 & = \cE Z_5 - \frac{F_4 F_1}{\cE} - \frac{F_3 F_2}{\cE} - 
          \frac{F_3 F_1^2}{2\cE^2} - \frac{F_2^2 F_1}{2\cE^2} -
          \frac{F_2 F_1^3}{6\cE^3} - \frac{F_1^5}{120\cE^4}~,\\
  \ldots & \ldots~.
 \end{aligned}
\end{equation}
The prepotential $F^{\mathrm{(n.p.)}}(\Phi)$ must be well-defined when the
closed string deformation is turned off, and hence all coefficients $F_k$ must
be finite in the limit $f\to 0$. On the other hand, as is clear from the
explicit expressions obtained in the previous section, the partition functions
$Z_k$ exhibit singularities of different orders, ranging from $O(f^{-4})$
(corresponding to $1/\cE$) up to $O(f^{-4k})$ (corresponding to $1/\cE^k$).
Thus, for consistency of the whole procedure, in computing $F_k$ all such
divergences must disappear. Imposing the cancellation of the most divergent term
fixes the overall normalization coefficients $\cN_k$ but, once this choice is
made, all the remaining cancellations of divergences must take place.

For $k=1$, from \eq{Z1fin} we have directly 
\begin{equation}
 \label{F1is}
  F_1 = \cN_1\,\Pf\phi~.
\end{equation}
For $k=2$, we must insert the above result into \eq{Fkexp} and use the
expression (\ref{Z2ris}) for the partition function $Z_2$. The resulting
contribution is
\begin{equation}
 \label{mdivF2}
 F_2 = \left(\frac{\cN_2}{4} - \frac{\cN_1^2}{2}\right)\frac{(\Pf\phi)^2}{\cE} 
+ \ldots~.
\end{equation}
We fix the normalization $\cN_2$ as
\begin{equation}
 \label{n2sol}
 \cN_2 = 2 \cN_1^2~
\end{equation}
in order to cancel the most divergent term, and having done so, we find
that all other divergences disappear, leaving 
\begin{equation}
 \label{F2sol}
 F_2= 2 \cN_1^2 \left(\frac{\Tr\phi^4 - \frac 12
(\Tr\phi^2)^2}{256}  + 
\frac{\Tr f^2\, \Tr \phi^2}{2048} + \frac{\Tr f^4 - \frac 54 (\Tr
f^2)^2}{16384}\right)~. 
\end{equation}
We proceed in the same way at the next order, $k=3$. Using \eq{solZ3} and the
above expressions for $F_1$ and $F_2$, one can see from \eq{Fkexp} that
the most divergent term of $F_3$ reads
\begin{equation}
\label{F3div}
F_3  = \frac{(\Pf\phi)^3}{\mathcal{E}^2} \left(\frac{\cN_3}{12} - \frac{\cN_2
\cN_1}{4} + \frac{\cN_1^3 }{3}\right) +  \ldots \\ 
= \frac{(\Pf\phi)^3}{\mathcal{E}^2} \left(\frac{\cN_3}{12} - \frac{\cN_1^3}{6}
\right) + \ldots~,
\end{equation}
so that we have to choose
\begin{equation}
 \label{n3sol}
 \cN_3 = 2 \cN_1^3~.
\end{equation}
Once this is done, all other divergences cancel and we are simply left with
\begin{equation}
 F_3 = \frac{\cN_1^3}{48} \,\Pf\phi~.
\label{F3is}
\end{equation}
It is interesting to note that the contributions 
from odd instanton numbers $k=2n+1$ have to contain the factor $\Pf\phi$ 
which, being quartic, saturates already the dimensionality of the prepotential.
Thus, in these cases, there is no room for $f$-dependent terms.

So far, the only ambiguity left is the overall normalization factor
$\cN_1$.
Considering the ratio $F_3/F_1={\cN_1^2}/{48}$, we see that by setting
\begin{equation}
\label{N1}
\cN_1 = 8~,
\end{equation}
it takes the value $d_3/d_1=4/3$ as in the heterotic theory (see \eq{exphetc}).
With this choice all possible ambiguities are fixed, and no further adjustments
are possible.
For $k=4$, the partition function $Z_4$ can be computed as indicated in Appendix
\ref{subapp:k45}. The cancellation of most divergent term in the expression of
$F_4$ following from \eq{Fkexp}, requires that $\cN_4 = 2 \cN_1^4 = 8192$. Using
this, we then find
\begin{equation}
 \label{f4res}
 F_4 = \frac 14 \Tr\phi^4 - \frac 14 (\Tr\phi^2)^2 + \frac{3}{32} \Tr\phi^2\, 
\Tr f^2 + \frac{3}{256} \left(\Tr f^4 - \frac 54 (\Tr f^2)^2\right)~.
\end{equation}
In the case $k=5$, having computed $Z_5$ along the lines described in Appendix
\ref{subapp:k45}, the cancellation of the highest divergence in $F_5$ requires
that $\cN_5 = 2 \cN_1^5 = 65536$, after which we get 
\begin{equation}
 F_5 = \frac{48}{5}\,\Pf\phi~.
\end{equation}

Making the replacement $\phi\to\Phi(x,\theta)$ and taking the limit $f\to 0$ in
the above results, we obtain the non-perturbative contributions to the
prepotential according to \eq{Fexp}. Up to instanton number $k=5$, our findings
are summarized in
\begin{equation}
 \label{Fupto5}
  \begin{aligned}
   F^{\mathrm{(n.p.)}}(\Phi)  = &~
\Tr\Phi^4\, \Big(\frac 12\, q^2 + \frac 14\, q^4 + \ldots\Big) - 
   (\Tr\Phi^2)^2\,\Big(\frac 14\, q^2 + \frac 14\, q^4 + \ldots\Big)\\
   &~ + 8\,\Pf\Phi\, \Big(q + \frac 43\, q^3 + \frac 65 \,q^5 + \ldots\Big)~,
  \end{aligned}
\end{equation}
which perfectly match the expectations from the heterotic string, as one can see
by comparing Eq.s~(\ref{Fupto5}) and (\ref{exphet})%
\footnote{The $t_8$ structure is produced, according to Eq.s~(\ref{R+-}) and
(\ref{inttheta}), by integrating the prepotential over $d^8\theta$ to obtain the
effective Lagrangian.}.

Actually, if we refrain from taking the limit $f\to 0$, our method allows to
obtain also the instanton-induced gravitational corrections to the prepotential.
Indeed, once the factor $1/\cE$ is removed from $\log\mathcal Z$ as indicated in
(\ref{prep1}), we are allowed not only to replace $\phi$ with the full dynamical
gauge superfield $\Phi(x,\theta)$ as we have done so far, but also to replace
the constant RR background $f$ with a full-fledged dynamical gravitational
superfield, in complete analogy with what happens in the $\mathcal N=2$ SYM
theories in four dimensions \cite{Billo:2006jm}. The reason is that the
$(7\times 7)$ matrix $f_{mn}$ defines, through \eq{Ff}, an $(8\times 8)$
anti-symmetric tensor $\cF_{\mu\nu}$, which can be interpreted as the
graviphoton field-strength. In turn, $\cF_{\mu\nu}$ can be considered as the
lowest component of a eight-dimensional bulk chiral superfield $\mathcal
W_{\mu\nu}(x,\theta)$ defined as
\begin{equation}
\label{grav}
\mathcal {W}_{\mu\nu}(x,\theta) = \mathcal{F}_{\mu\nu}(x) + \theta\,\chi_{\mu\nu}+
\frac{1}{2}\,
\theta \gamma^{\rho\sigma} \theta \,{\mathcal R}_{\rho\sigma\mu\nu}(x) + \cdots
\end{equation}
where $\chi_{\mu\nu}$ is the gravitino field-strength and $\mathcal
R_{\rho\sigma\mu\nu}$ is the Riemann curvature tensor. Notice that since the
matrix $f_{mn}$ parameterizes the 21 components of $\mathcal{F}_{\mu\nu}$ that
are related to the rotation in a 7-dimensional subspace as indicated in \eq{Ff},
the graviphoton field-strength is subject to the constraint
\begin{equation}
\label{p1}
(P_1^+)^{\mu\nu}_{\rho\sigma}\, {\cal F}_{\mu\nu} =0~,
\end{equation}
where $P_1^+$ is the octonionic projector described in appendix
\ref{subapp:octo}. This constraint can be viewed as the eight-dimensional
analogue of the self-duality constraint that is imposed on the graviphoton
background in four dimensions \cite{Nekrasov:2002qd,Billo:2006jm}.

In this way we can obtain the non-perturbative prepotential, including
gravitational corrections, which is therefore given by 
\begin{equation}
 \label{prep2}
 F^{\mathrm{(n.p.)}}(\Phi,\mathcal W) = \mathcal{E}\, \log 
\mathcal Z\Big|_{\phi\to\Phi,f\to \mathcal W}~.
\end{equation}
The first few contributions at low instanton numbers can be read from
Eq.s~(\ref{F1is}), (\ref{F2sol}) and (\ref{f4res}). However, to express the
result in a covariant form, it is convenient to first take advantage of the
following trace identities
\begin{equation}
\label{fF}
\begin{aligned}
\Tr f^2  &= \frac{1}{4}\, \Tr {\cF}^2 ~,\\
\Tr f^4 - \frac{5}{4} (\Tr f^2)^2 & = -\frac{1}{8}
\Big( \Tr {\cF}^4 +\frac14 (\Tr {\cF}^2)^2 \Big)~,
\end{aligned}
\end{equation}
so that we obtain
\begin{equation}
\begin{aligned}
\label{gravris}
F^{\mathrm{(n.p.)}}(\Phi,\mathcal W) = & ~F^{\mathrm{(n.p.)}}(\Phi) +
\frac{1}{2^6}\,\Tr{\mathcal W}^2\, \Tr \Phi^2\,
\Big(q^2+\frac{3}{2}\,q^4+\cdots\Big)\\
&-\frac{1}{2^{10}}\Big( \Tr {\mathcal W}^4 + \frac 14 (\Tr
{\mathcal W}^2)^2\Big)\,\Big(q^2+\frac{3}{2}\,q^4+\cdots\Big)~.
\end{aligned}
\end{equation}
Once we perform the integration over the fermionic superspace coordinates, this
expression shows that  instantons with even topological charge induce in the
D7-brane effective action non-perturbative purely gravitational terms
proportional to $\Tr {\mathcal R}^4 + \frac14 (\Tr {\mathcal R}^2)^2$, and mixed
gauge/gravitational terms proportional to $\Tr {\mathcal R}^2\,\Tr {F}^2$. The
relative coefficients of the instanton corrections for the various structures
are again in perfect agreement with the expectations from the heterotic string
calculations, as indicated in \eq{gravhet}.

\section{Conclusions}
\label{sec:concl}
In this paper we have analyzed in detail the integral over the D-instanton
moduli in the type I$^\prime$ theory. Such matrix integrals are different from
the D(--1) matrix integrals in type IIB since they possess mixed moduli from the
D7/D(--1) sectors. They also differ from ``ordinary'' instantonic brane systems,
such as the D3/D(--1) system, because the mixed moduli are only fermionic; they
are instead similar to so-called ``exotic'' instantons. We have shown that
localization techniques similar to the ones that were successful for type IIB
matrix integrals and for the $\cN=2$ instanton calculus in four dimensions allow
to perform the integration also in the D7/D(-1) system for generic values of the
instanton number $k$.  The outcome of the computation is the quartic
prepotential for the SO$(8)$ gauge multiplet $\Phi(x,\theta)$ on a stack of
D7-branes. Up to $k=5$ and taking into account also the tree-level and one-loop
contributions discussed in Section \ref{subsec:gauge}, the explicit result we
find is
\begin{equation}
 \label{resprep}
 \begin{aligned}
   F(\Phi) = & ~\Tr \Phi^4 \Big[\frac{\ii\pi\tau}{12} + \frac 12 q^2 +
   \frac 14 q^4 + \ldots\Big] \\
   & + \big(\Tr\Phi^2\big)^2 \Big[\frac{1}{32} \log\big(\im\tau\,\im 
   U \left|\eta(U)\right|^4\big)- \frac 14 q^2 - \frac 14 q^4 + \ldots
   \Big] \\
   & + 8\, \Pf\Phi \Big[q + \frac 43 q^3 + \frac 65 q^5 + \ldots\Big]~.
 \end{aligned}
\end{equation}
Using the duality relations (\ref{dualrel}) this expression matches perfectly
the results of the heterotic $\soq$ theory.
Our computation represents thus an explicit quantitative check of the
heterotic/type I$^\prime$ duality. 

We expect that the techniques we utilized may be useful in dealing with the
moduli space integrals for other instances of ``exotic'' instanton systems,
also the four-dimensional ones of potential phenomenological relevance.

\vskip 1cm
\noindent {\large {\bf Acknowledgments}}
\vskip 0.2cm
\noindent We would like to thank C. Bachas, G. Ferretti, F. Fucito, J. F.
Morales and R. Poghossian for several illuminating discussions. A.L. thanks the
Galileo Galilei Institute for Theoretical Physics for the hospitality and the
INFN for partial support during the completion of this work. This research was
supported in part by the French Agence Nationale de la Recherche under grant
ANR-06-BLAN-0142.

\appendix

\section{Conventions and notations}
\label{appa}

\subsection{SO$(7)$ and SO$(8)$ gamma-matrices}
\label{subapp:octo}

SO$(7)$ can be embedded into SO$(8)$ in such a way that the vector
representation $\mathbf{8}$ of SO$(8)$ is identified with  the spinor
representation $\mathbf{8}_s$ of SO$(7)$. This embedding is best described by
using an explicit realization of the Clifford algebras in $d=7$ and $d=8$ based
on the octonionic structure constants.

The Clifford algebra  in 7 dimensions,
\begin{equation}
\label{cl7tau}
\acomm{\tau^i}{\tau^j}_{\alpha\beta} = -2\delta^{ij}\,\delta_{\alpha\beta}~,
\end{equation}
can be realized by the matrices $(8\times 8)$-matrices $\tau^i$ ($i=1,\ldots,7$)
with elements 
\begin{equation}
(\tau^i)_{\alpha\beta} =
 \delta^{i8}_{\alpha\beta}+C^{-\,i8}_{\alpha\beta}\quad\quad
(\alpha,\beta=1,\ldots, 8)~,
 \label{gamma7}
\end{equation}
where we made use of the totally antisymmetric, (anti)-selfdual four-index
tensors in $d=8$ $C_{\mu\nu\rho\sigma}^\pm$. In turn, these tensors are
expressed as
\begin{equation}
C_{ijk8}^\pm = c_{ijk} \quad,\quad
C_{ijk\ell}^\pm = \pm \frac{1}{3!} \,\epsilon_{ijk\ell mnp}\,c_{mnp}
\label{Cpm}
\end{equation}
in terms of the octonionic structure constants $c_{mnp}$ ($m,n,\ldots =
1,\ldots, 7$), with $c$ a totally antisymmetric tensor whose only non-zero
elements can be taken to be
\begin{equation}
c_{127}=c_{163}=c_{154}=c_{253}=c_{246}=c_{347}=c_{567}=1~.
\label{c}
\end{equation}
The tensor $c_{mnp}$ enjoys various properties, such as 
\begin{equation}
\label{propc}
c_{mpr}c_{nrq} = -\delta_{mn}\delta_{pq} + \delta_{mq}\delta_{np} + 
\frac{1}{6} \epsilon_{mnpqstu}c_{stu}
\equiv -\delta_{mp,nq} + \frac{1}{6} \epsilon_{mnpqstu}c_{stu}
\end{equation}
and
\begin{equation}
\label{propc2}
c_{mpq}c_{npq}= 6\, \delta_{mn}~,~~~ 
\epsilon_{mnpqstu}c_{stu}c_{rpq} = 24\, c_{mnr}~.
\end{equation}
These properties imply the existence of useful identities for the tensors
$C_{\mu\nu\rho\sigma}^\pm$, such as 
\begin{equation}
 \label{idC}
 C^{\pm\mu\nu\rho\sigma}C^{\pm}_{\rho\sigma\tau\omega} = 
6\, \delta^{\mu\nu}_{\tau\omega} \pm 4\,  C^{\mu\nu}_{\tau\omega}~.
\end{equation}

The SO$(7)$ generators $T^{ij}$ in the spinorial representation $\mathbf{8}_s$,
satisfying the $\mathrm{so}(7)$ algebra 
\begin{equation}
 \bigcomm{T^{mn}}{T^{pq}} = \delta^{np}\,T^{mq} - \delta^{mp}\,T^{nq}
+ \delta^{mq}\,T^{np}-\delta^{nq}\,T^{mp} ~,
\label{TT}
\end{equation}
are defined as
\begin{equation}
\label{defT}
 T^{mn}= -\frac{1}{4}\,\comm{\tau^m}{\tau^n}\equiv -\frac{1}{2}\,\tau^{mn}~.
\end{equation}
Using the definition (\ref{gamma7}) and \eq{idC}, one can show that 
\begin{equation}
 \label{expT}
 \big(T^{mn}\big)_{AB}= \frac{1}{2}\Big(\delta^{mn}_{\,AB}+ C^{-\,mn}_{~AB}\Big)~.
\end{equation}
A generic $\mathrm{SO}(7)$ group element in the spinor representation can then
be parametrized as
\begin{equation}
 R(f) = \ee^{\frac{1}{2}f_{mn}\,T^{mn}}
\label{R}
\end{equation}
and the infinitesimal $\mathrm{SO}(7)$ variation of any field $X^A$ transforming
in the spinor representation is 
\begin{equation}
 \delta X^A = \frac{1}{2}\,f_{mn}\big(T^{mn}\big)^{AB}\,X_B = -\frac{1}{2}\,
\mathcal{F}^{AB}_{\mathbf{21}}\,X_B
\end{equation}
where we have introduced (the subscript $\mathbf{21}$ will become clear later)
\begin{equation}
 \mathcal{F}^{AB}_{\mathbf{21}}= \frac{1}{2}\,f_{mn}\big(\tau^{mn}\big)^{AB}~.
\end{equation}

The SO$(7)$ generators $t^{mn}$ in the vector representation $\mathbf{7}$,
satisfying the $\mathrm{so}(7)$ algebra with the \emph{same} normalization as in
\eq{TT} are given by
\begin{equation}
 \label{deft}
 \big(t^{mn}\big)_{pq} = \delta^{mn}_{pq} = 
\delta^m_p\, \delta^n_q - \delta^m_q\, \delta^n_p~.
\end{equation}
Thus, in the vector representation, the $\mathrm{SO}(7)$ group element with
parameters $f_{mn}$ is represented by 
\begin{equation}
 r(f) = \ee^{\frac{1}{2}f_{mn}\,t^{mn}}
\label{r}
\end{equation}
and the infinitesimal variation of any field $\phi^m$ transforming in the vector
representation is 
\begin{equation}
 \delta \phi^p = \frac{1}{2}\,f_{mn}\big(t^{mn}\big)^{pq}\phi_q =
\frac{1}{2}\,f_{mn}\,\delta^{mn,pq}\,\phi_q= f^{pq}\,\phi_q~.
\end{equation}

The SO$(8)$ Clifford algebra can be realized by taking the eight
gamma matrices $\gamma^\mu$ to be (we use now $\mu=1,\ldots 8$, while 
$m=1,\ldots 7$)
\begin{equation}
\label{so8oct}
\gamma^{m} = \ii\tau^{m} \otimes \sigma^1~,~~~
\gamma^{8}=  \mathbf{1}_{8} \otimes \sigma^2~.
\end{equation}
These matrices satisfy indeed
\begin{equation}
\comm{\gamma^{\mu}}{\gamma^{\nu}} = 2 \delta^{\mu\nu}~.
\end{equation}
Note that this is a Weyl basis, since the chirality matrix $\gamma$ is
represented by
\begin{equation}
 \label{chiralgammaoct}
\gamma \equiv - \gamma^1 \gamma^2 \ldots \gamma^8 = - \tau^1\tau^2\ldots\tau^7
\otimes \sigma^3
= \mathbf{1} \otimes \sigma^3~.
\end{equation}
Note also that in their realization given in \eq{so8oct} all the gamma matrices
are antisymmetric; the charge conjugation matrix can thus be taken to
be simply the identity matrix. 

The two-index gamma-matrices 
\begin{equation}
 \label{defgammamunu}
 \gamma^{\mu\nu}\equiv \frac 12 \comm{\gamma^\mu}{\gamma^\nu}~,
\end{equation}
which are again anti-symmetric, are given, according to \eq{so8oct}, by
\begin{equation}
\label{commgamma}
\gamma^{mn} =  -\tau^{mn} \otimes \mathbf{1}_{2}~,~~~
\gamma^{m8}=  -\tau^{m}\otimes \sigma^3~.
\end{equation}
For the anti-chiral block we find explicitly
\begin{equation}
\label{commgamma2a}
\bar\gamma^{\mu\nu}_{\dot\alpha \dot\beta} = 
C^{-\mu\nu}_{\dot\alpha\dot\beta} +\delta^{\mu\nu}_{\dot\alpha \dot\beta}
\end{equation}
while for the chiral block we can write
\begin{equation}
\label{commgamma2c}
(\gamma^{mn})_{\alpha\beta} = \delta^{mn}_{\alpha\beta} +
C^{-mn}_{\alpha\beta}~,
~~~ (\gamma^{m8})_{\alpha\beta} = -\delta^{m8}_{\alpha\beta}
-C^{-m8}_{\alpha\beta}
\end{equation}
or, splitting the spinor index $\alpha$ into $(a,8)$ with $a=1,\ldots 7$,
\begin{equation}
\label{commgamma2cbis}
(\gamma^{\mu\nu})_{ab} = \delta^{\mu\nu}_{ab} - C^{+\mu\nu}_{ab}~,~~~
(\gamma^{\mu\nu})_{a8} = -\delta^{\mu\nu}_{a8} + C^{+\mu\nu}_{a8}~.
\end{equation}

The 28-dimensional space of anti-symmetric $8\times 8$ matrices, namely the
adjoint space of SO$(8)$, admits an orthogonal decomposition $\mathbf{28}\to
\mathbf{21} + \mathbf{7}$ enforced by the following projectors:
\begin{equation}
\begin{aligned}
(P_1^+)^{\mu\nu}_{\rho\sigma} & = 
\frac 18\left(\delta^{\mu\nu}_{\rho\sigma} +
C^{+\mu\nu}_{~~~\rho\sigma}\right)~,\\
(P_2^+)^{\mu\nu}_{\rho\sigma} & = 
\frac 38\left(\delta^{\mu\nu}_{\rho\sigma} -\frac 13
C^{+\mu\nu}_{~~~\rho\sigma}\right)~.
\end{aligned}
\end{equation}
Indeed, it is straightforward to check that 
\begin{equation}
\label{pro12} 
(P_1^+)^2 = P_1^+~,~~~
(P_2^+)^2 = P_2^+~,~~~ 
P_1^+ P_2^+ = P_2^+ P_1^+ = 0~,~~~ 
P_1^+ + P_2^+ = \mathbf{1}
\end{equation}
using the properties of the tensor $C^+_{\mu\nu\rho\sigma}$, see \eq{idC}. Since
the tensor $C^+_{\mu\nu\rho\sigma}$ is traceless, the dimensionality of the two
eigenspaces are easily obtained by taking the trace of the projectors:
\begin{equation}
 \label{kers}
\dim\, \mathrm{Ker}(P_1^+) = 21~,~~~
\dim\, \mathrm{Ker}(P_2^+) = 7~.
\end{equation}

The $\mathrm{Ker}(P_1^+)$ subspace is spanned by the 21 matrices
$(\tau^{mn})_{\mu\nu}$ corresponding to (twice) the SO$(7)$ spinorial generators
in which we identify the indices $A$ in the $\mathbf{8}_s$ of SO$(7)$ with the
indices $\mu$ in the vector of SO$(8)$. Indeed one can verify that
\begin{equation} 
\label{verproj1}
(P_1^+)^{\mu\nu}_{\rho\sigma} (\tau^{mn})_{\mu\nu} \propto 
\left(\delta^{\mu\nu}_{\rho\sigma} + 
C^{+\mu\nu}_{~~~\rho\sigma}\right) (\tau^{mn})_{\mu\nu}  = 0~.
\end{equation}
The $\mathrm{Ker}(P_2^+)$ subspace is instead spanned by the 7 matrices
$(\tau^{m})_{\mu\nu}$, namely the SO$(7)$ matrices with the above identification
of spinorial indices of SO$(7)$ and vector indices of SO$(8)$: \begin{equation}
\label{verproj2}
(P_2^+)^{\mu\nu}_{\rho\sigma} (\tau^{m})_{\mu\nu} \propto 
\left(\delta^{\mu\nu}_{\rho\sigma} -
\frac 13 C^{+\mu\nu}_{~~~\rho\sigma}\right)(\tau^{m})_{\mu\nu}  = 0~.
\end{equation}

Thus, there is a non-standard%
\footnote{In a standard embedding, the adjoint
representation $\mathbf{21}$ of SO$(7)$ corresponds simply to the restriction of
$\mathcal{F}_{\mu\nu}$ to its elements $\mathcal{F}_{mn}$, while the
$\mathbf{7}$ corresponds to $\mathcal{F}_{m8}$. } embedding of SO$(7)$ into
SO$(8)$ in which the adjoint representation of the latter, whose elements are
antisymmetric $8\times 8$ matrices $\mathcal{F}_{\mu\nu}$, decomposes into
$\mathbf{21} \oplus \mathbf{7}$ as follows:
\begin{equation}
 \label{deco28}
\mathcal{F}_{\mu\nu} = \mathcal{F}_{\mu\nu}^{\mathbf{21}} + 
\mathcal{F}_{\mu\nu}^{\mathbf{7}} =
\frac 12 f_{mn} (\tau^{mn})_{\mu\nu} + h_m (\tau^m)_{\mu\nu}~.
\end{equation}

Eq.s (\ref{verproj1}) and (\ref{verproj2}) imply the following relations, useful
in the computation of the diagram in Fig. \ref{fig2} a): 
\begin{equation}
\begin{aligned}
 (\tau^{mn})_{\mu\nu}\,\left(\delta^{\mu\nu}_{\rho\sigma} - 
C^{+\mu\nu}_{~~~\rho\sigma}\right) & =+ \,4 \,(\tau^{mn})_{\rho\sigma}~,\\
 (\tau^{m})_{\mu\nu}\,\left(\delta^{\mu\nu}_{\rho\sigma} -
C^{+\mu\nu}_{~~~\rho\sigma}\right) & =- \,4 \,(\tau^{m})_{\rho\sigma}~.
\end{aligned}
\label{tauC}
\end{equation}

The following identities are instead useful for the computation of the diagram
in Fig. \ref{fig1} a): 
\begin{equation}
\label{idtau2gamma2}
\begin{aligned}
(\tau^{mn})_{\mu\nu} (\bar\gamma^{\mu\nu})_{pq} &= 
- 8 \delta^{mn}_{pq}\quad,\quad
(\tau^{mn})_{\mu\nu} (\bar\gamma^{\mu\nu})_{p8} = 0~, \\
(\tau^{m})_{\mu\nu} (\bar\gamma^{\mu\nu})_{p8} &= +8 \delta^{m~}_{p~}\quad,\quad
(\tau^{m})_{\mu\nu} (\bar\gamma^{\mu\nu})_{pq} = 0~.
\end{aligned}
\end{equation}

\subsection{The $t_8$ tensor}
\label{subapp:t8}
The explicit expression of the totally anti-symmetric 8-index tensor $t_8$ can
be read from \eq{f4}. Several of its properties are given, for instance, in
Appendix B of \cite{Billo':2009gc}. Here, let us just recall how it appears from
the integration over the superspace coordinates $d^8\theta$ (or $d^8\bar\theta$)
of chiral (or anti-chiral) superfields such as  those in \eq{Phi} or  \eq{grav},
see for example Appendix 9.A of Ref. \cite{Green:1987mn}. For bi-linear
operators of the form
\begin{equation}
 R^{\mu\nu} = \frac{1}{4} (\gamma^{\mu\nu})_{\alpha\beta} \theta^\alpha 
\theta^\beta~,~~~
\bar R^{\mu\nu} = \frac{1}{4} (\bar\gamma^{\mu\nu})_{\dot\alpha\dot\beta} 
{\bar\theta}^{\dot\alpha} {\bar\theta}^{\dot\beta}~,
\label{R+-}
\end{equation}
one finds 
\begin{equation}
  \label{inttheta}
  \int d^8\theta \big(R^{\mu_1\mu_2}\cdots R^{\mu_7\mu_8}\big) = 
  t_+^{\mu_1\mu_2\ldots \mu_7\mu_8}~,~~
  \int d^8\bar\theta \big(\bar R^{\mu_1\mu_2} \cdots \bar R^{\mu_7\mu_8}\big) = 
  t_-^{\mu_1\mu_2\ldots \mu_7\mu_8}
\end{equation}
with the antisymmetric tensors $t_\pm$ being related to $t_8$ and to the
Levi-Civita tensor $\epsilon_8$ by 
\begin{equation}
 \label{tpmt8}
 t_\pm = t_8 \pm \frac 12 \epsilon_8~.
\end{equation}

\section{Vertex operators and disk amplitudes}
\label{appb}

In this Appendix we give some details on the evaluation of the string diagrams
that describe the interaction 
between the instanton moduli and the constant RR background.

As explained in Ref.s \cite {Billo:2002hm,Billo:2004zq,Billo:2006jm}, to
simplify the procedure it is convenient to first rewrite the quartic
interactions among $a_\mu$, $\chi$ and $\bar \chi$ appearing in ${\mathcal
S'}_{\mathrm{quartic}}$ of \eq{quartic} in a cubic form.
This can be done by introducing two new auxiliary fields $Y_\mu$ and $\bar
Y^\mu$, so that
we can replace ${\mathcal S'}_{\mathrm{quartic}}$ by 
\begin{equation}
\label{saux3}
\begin{aligned}
{\mathcal S'}_{\mathrm{aux}} = & \frac{1}{g_{0}^2}\, \tr
\Big\{
\frac{1}{2}D_m D^m + \frac{1}{2}D_m (\tau^m)_{\mu\nu} \comm{a^{\mu}}{a^{\nu}}
 \\
&\hspace{20pt}
+ \bar{Y}^{\mu} Y_{\mu}+ \left[a^\mu,\bar\chi\right]Y_{\mu}
+\bar{Y}^{\mu}\left[a_\mu,\chi\right]
+\frac{1}{2} \comm{\bar\chi}{\chi}^2
\Big\}~.
\end{aligned}
\end{equation}
It is easy to see that ${\mathcal S'}_{\mathrm{aux}}$ reduces to ${\mathcal
S'}_{\mathrm{quartic}}$ when the auxiliary fields $Y_\mu$ and $\bar Y^\mu$
acquire their on-shell values:
\begin{equation}
\label{auxval}
Y_{\mu} =-\left[a_{\mu},\chi\right]~,~~~
\bar{Y}^{\mu}=-\left[a^\mu,\bar\chi\right]~.
\end{equation}
The entire moduli action ${\mathcal S'}_{\mathrm{cubic}}+{\mathcal
S'}_{\mathrm{mixed}}+ {\mathcal S'}_{\mathrm{aux}}$ (with the first two terms
given in Eq.s (\ref{cubic1}) and (\ref{mixed1})) can be obtained by computing
``scattering'' amplitudes among the vertex operators representing the various
instanton moduli, including the auxiliary ones. In standard CFT notations (see
for example Ref.s~\cite{Billo:2002hm,Billo':2009gc} for details), these vertex
operators are
\begin{equation}
\label{vertexNS}
V_a =  (2 \pi \alpha')^{1/2} \,a_\mu \, \psi^\mu\,\ee^{-\varphi}~,~~
V_\chi = (2 \pi \alpha')^{1/2}\, \chi \, {\bar \Psi}\,\ee^{-\varphi}~,~~  
V_{\bar \chi} = (2 \pi \alpha')^{1/2}\,{\bar \chi}\,  \Psi\,\ee^{-\varphi}~,
\end{equation}
for the neutral moduli of the Neveu-Schwarz sector, and
\begin{equation}
\label{vertexR}
V_M = (2 \pi \alpha')^{3/4}\, M_{\alpha}\,  S^{\alpha}\,S^- \,\ee^{-\varphi/2}~,
~~
V_{\lambda} = (2 \pi \alpha')^{3/4}\, \lambda_{\dot \alpha}\, 
S^{\dot \alpha}\,S^+\,\ee^{-\varphi/2}~,
\end{equation}
for those of the Ramond sector. For the fermionic charged moduli, corresponding
to open strings with eight mixed ND directions, we have instead
\begin{equation}
 \label{vertexmu}
V_\mu = (2 \pi \alpha')^{3/4}\, \mu\,  \Delta\,S^+ \,\ee^{-\varphi/2}~.
\end{equation}
Finally, the vertex operators for the auxiliary moduli are 
\begin{equation}
V_D = (2\pi \alpha')\, D_m (\tau^m)_{\mu\nu}\, :\psi^\mu\psi^\nu:~,~~
V_Y =(2 \pi \alpha')\,Y_\mu \, :{\bar \Psi} \psi^\mu: ~,~~
V_{\bar Y} =(2 \pi \alpha')\,{\bar Y}_\mu \, :\Psi\psi^\mu:
\label{vertaux}
\end{equation}
in the neutral sector, and
\begin{equation}
V_w = (2\pi \alpha')\, w\,\Delta\,S^{\dot\alpha=\dot 8}
\label{vertaux1}
\end{equation}
in the charged sector. In writing these vertex operators, we have neglected all
numerical factors in the normalizations and only inserted the appropriate powers
of $(2\pi\alpha')$ that are needed to give the moduli the canonical dimensions
(not the ADHM ones). Indeed, as we have shown in the main text, the result of
the integration over the moduli space is insensitive to the numerical
coefficients of the various structures. 

Notice that in \eq{vertaux1} we have selected the $\dot\alpha=\dot 8$ component
of the spin field $S^{\dot\alpha}$, since the BRST charge used in Section
\ref{sec:BRSmod} is precisely the $\dot\alpha=\dot 8$ component of the
supersymmetry charge $Q^{\dot\alpha}$, which is preserved by both the D7- and
the D(--1)-branes, and given by
\begin{equation}
\label{qal}
Q^{\dot \alpha} = \oint \frac{dz}{2\pi \ii} \, S^{\dot \alpha}(z)\,S^+(z) \ee^{-\varphi(z)/2}~.
\end{equation}
Using this information, and applying the techniques discussed in
Ref.s~\cite{Green:2000ke,Billo:2002hm}, one can check the BRST transformation
properties reported in \eq{Q}, as well as
\begin{equation}\
\label{Qaux}
Q M^\mu= \ii \sqrt{2}\, Y^\mu~, ~~~QY^\mu = - [M^\mu, \chi] ~.
\end{equation}

\begin{figure}[htb]
 \begin{center}
  \begin{picture}(0,0)%
\includegraphics{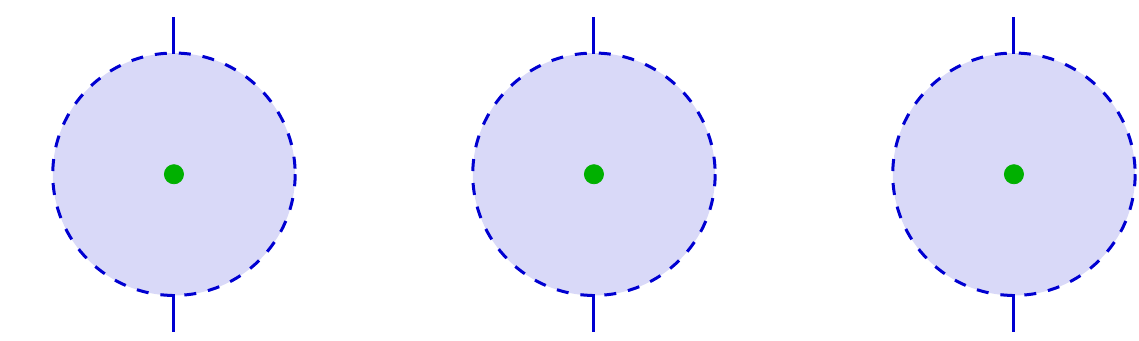}%
\end{picture}%
\setlength{\unitlength}{1895sp}%
\begingroup\makeatletter\ifx\SetFigFontNFSS\undefined%
\gdef\SetFigFontNFSS#1#2#3#4#5{%
  \reset@font\fontsize{#1}{#2pt}%
  \fontfamily{#3}\fontseries{#4}\fontshape{#5}%
  \selectfont}%
\fi\endgroup%
\begin{picture}(11375,3381)(136,-2923)
\put(4351,239){\makebox(0,0)[lb]{\smash{{\SetFigFontNFSS{9}{10.8}{\familydefault}{\mddefault}{\updefault}b)}}}}
\put(8551,239){\makebox(0,0)[lb]{\smash{{\SetFigFontNFSS{9}{10.8}{\familydefault}{\mddefault}{\updefault}c)}}}}
\put(6226,-2836){\makebox(0,0)[lb]{\smash{{\SetFigFontNFSS{9}{10.8}{\familydefault}{\mddefault}{\updefault}$\bar\chi$}}}}
\put(6226,164){\makebox(0,0)[lb]{\smash{{\SetFigFontNFSS{9}{10.8}{\familydefault}{\mddefault}{\updefault}$D$}}}}
\put(5926,-1036){\makebox(0,0)[lb]{\smash{{\SetFigFontNFSS{9}{10.8}{\familydefault}{\mddefault}{\updefault}$\cF$}}}}
\put(10426,-2836){\makebox(0,0)[lb]{\smash{{\SetFigFontNFSS{9}{10.8}{\familydefault}{\mddefault}{\updefault}$a$}}}}
\put(10426,164){\makebox(0,0)[lb]{\smash{{\SetFigFontNFSS{9}{10.8}{\familydefault}{\mddefault}{\updefault}$\bar Y$}}}}
\put(10126,-1036){\makebox(0,0)[lb]{\smash{{\SetFigFontNFSS{9}{10.8}{\familydefault}{\mddefault}{\updefault}$\cF$}}}}
\put(151,239){\makebox(0,0)[lb]{\smash{{\SetFigFontNFSS{9}{10.8}{\familydefault}{\mddefault}{\updefault}a)}}}}
\put(2026,-2836){\makebox(0,0)[lb]{\smash{{\SetFigFontNFSS{9}{10.8}{\familydefault}{\mddefault}{\updefault}$\lambda$}}}}
\put(2026,164){\makebox(0,0)[lb]{\smash{{\SetFigFontNFSS{9}{10.8}{\familydefault}{\mddefault}{\updefault}$\lambda$}}}}
\put(1726,-1036){\makebox(0,0)[lb]{\smash{{\SetFigFontNFSS{9}{10.8}{\familydefault}{\mddefault}{\updefault}$\cF$}}}}
\end{picture}%
 \end{center}
 \caption{Disk diagrams describing the interactions of a holomorphic RR field-strength vertex 
 (in the interior of the disk) with moduli vertices. 
 The boundary of the disk is on the D(--1)'s.}
 \label{fig1}
\end{figure}
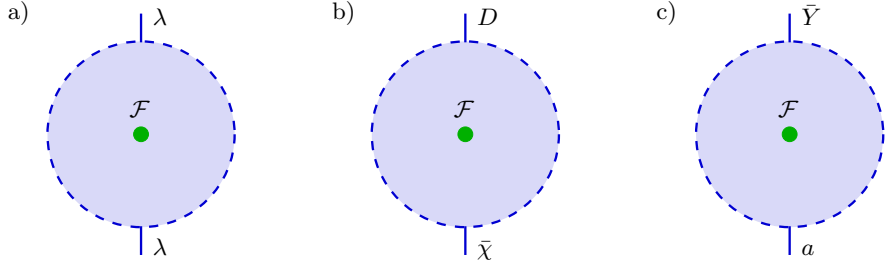

\begin{figure}[htb]
 \begin{center}
  \begin{picture}(0,0)%
\includegraphics{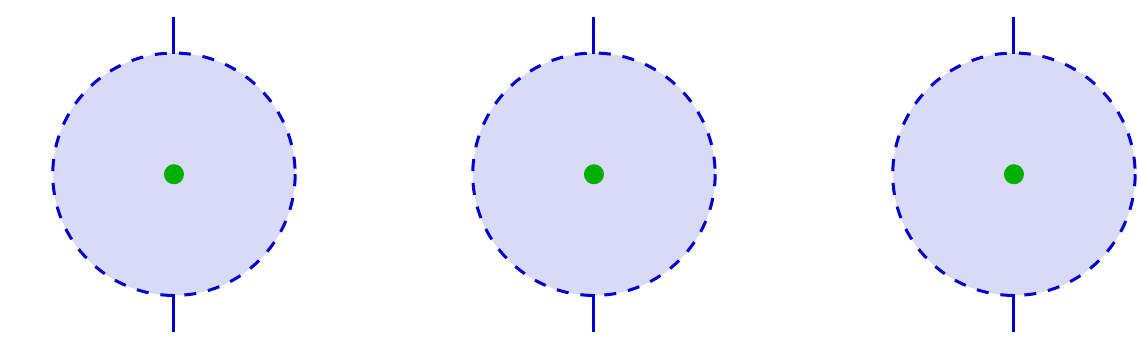}%
\end{picture}%
\setlength{\unitlength}{1895sp}%
\begingroup\makeatletter\ifx\SetFigFontNFSS\undefined%
\gdef\SetFigFontNFSS#1#2#3#4#5{%
  \reset@font\fontsize{#1}{#2pt}%
  \fontfamily{#3}\fontseries{#4}\fontshape{#5}%
  \selectfont}%
\fi\endgroup%
\begin{picture}(11375,3381)(136,-2923)
\put(4351,239){\makebox(0,0)[lb]{\smash{{\SetFigFontNFSS{9}{10.8}{\familydefault}{\mddefault}{\updefault}b)}}}}
\put(8551,239){\makebox(0,0)[lb]{\smash{{\SetFigFontNFSS{9}{10.8}{\familydefault}{\mddefault}{\updefault}c)}}}}
\put(6226,-2836){\makebox(0,0)[lb]{\smash{{\SetFigFontNFSS{9}{10.8}{\familydefault}{\mddefault}{\updefault}$\chi$}}}}
\put(6226,164){\makebox(0,0)[lb]{\smash{{\SetFigFontNFSS{9}{10.8}{\familydefault}{\mddefault}{\updefault}$D$}}}}
\put(5926,-1036){\makebox(0,0)[lb]{\smash{{\SetFigFontNFSS{9}{10.8}{\familydefault}{\mddefault}{\updefault}$\bcF$}}}}
\put(10426,-2836){\makebox(0,0)[lb]{\smash{{\SetFigFontNFSS{9}{10.8}{\familydefault}{\mddefault}{\updefault}$a$}}}}
\put(10426,164){\makebox(0,0)[lb]{\smash{{\SetFigFontNFSS{9}{10.8}{\familydefault}{\mddefault}{\updefault}$Y$}}}}
\put(10126,-1036){\makebox(0,0)[lb]{\smash{{\SetFigFontNFSS{9}{10.8}{\familydefault}{\mddefault}{\updefault}$\bcF$}}}}
\put(151,239){\makebox(0,0)[lb]{\smash{{\SetFigFontNFSS{9}{10.8}{\familydefault}{\mddefault}{\updefault}a)}}}}
\put(2026,-2836){\makebox(0,0)[lb]{\smash{{\SetFigFontNFSS{9}{10.8}{\familydefault}{\mddefault}{\updefault}$M$}}}}
\put(2026,164){\makebox(0,0)[lb]{\smash{{\SetFigFontNFSS{9}{10.8}{\familydefault}{\mddefault}{\updefault}$M$}}}}
\put(1726,-1036){\makebox(0,0)[lb]{\smash{{\SetFigFontNFSS{9}{10.8}{\familydefault}{\mddefault}{\updefault}$\bcF$}}}}
\end{picture}%
 \end{center}
 \caption{Disk diagrams describing the interactions of an anti-holomorphic RR
field-strength vertex (in the interior of the disk) with moduli vertices. 
 The boundary of the disk is on the D(-1)'s.}
 \label{fig2}
\end{figure}

This stringy approach to the instanton calculus allows to easily compute also
the interactions between moduli and bulk gravitational fields. In particular, we
are interested in the interactions with RR field-strengths $\mathcal F$ and
$\bar{\mathcal F}$, which correspond to the disk diagrams represented in Fig.s
\ref{fig1} and \ref{fig2}. 

These can be computed using standard CFT techniques
by inserting in the disk interior the following RR vertex operators
\begin{equation}
\label{vertexf}
\begin{aligned}
V_{\cal F} &= (2 \pi \alpha')^{1/2} \,{\cal F}_{\mu\nu} \,
(\gamma^{\mu\nu}\gamma)_{\alpha\beta}\,
S^{\alpha}\,S^-\,\ee^{-\varphi/2}~ {\widetilde S}^{\beta}\,
{\widetilde S}^-\,\ee^{-{\widetilde \varphi}/2}~,\\
V_{\bar {\cal F}} &= (2 \pi \alpha')^{1/2} \,{\bar {\cal F}}_{\mu\nu} 
\,(\gamma^{\mu\nu}\gamma)_{\dot\alpha \dot \beta}\,
S^{\dot \alpha}\,S^+\, \ee^{-\varphi/2}~ 
{\widetilde S}^{\dot \beta}\,{\widetilde S}^+\, \ee^{-{\widetilde \varphi}/2}~,
\end{aligned}
\end{equation}
where the matrices $\gamma^{\mu\nu}$ and $\gamma$ have been defined in
Eq.s~(\ref{defgammamunu}) and (\ref{chiralgammaoct}).

Let us now give some details on the computation of the disk diagram
represented in Fig. \ref{fig1}a, which corresponds to the following amplitude
\begin{equation}
\lvev V_{\lambda} V_{\lambda}V_{\cF} \rvev
\,\,\equiv\,\,
C_{0}\!
\int\frac{dx_1\, dx_2\, dz d\bar z}{dV_{\rm
CKG}}\,\,\times
\, \big\langle V_{\lambda}(x_1)\, V_{\lambda}(x_2)\,V_{\cF}(z,\bar z)\big\rangle
\end{equation}
where $C_{0}$ is the normalization of D(--1) disk amplitudes \cite {Billo:2002hm}
\begin{equation}
C_{0}= \frac{2}{(2 \pi\alpha')^2}\frac{1}{g_0^2}
\end{equation}
and $dV_{\mathrm{CKG}}$ is the volume of the conformal Killing group. As usual,
the open string punctures $x_i$ are integrated along the real axis with $x_1\geq
x_2$ while the closed string puncture $z$ is integrated on the upper half
complex plane. More explicitly, after reflecting the right movers on the disk
boundary, we have
\begin{equation}
\label{corr}
\begin{aligned}
& \lvev V_{\lambda} V_{\lambda}V_{\cF} \rvev
= \frac{2}{g_0^2}\,
{\tr}\Big(\lambda_{\dot\alpha}\lambda_{\dot\beta}\cF_{\mu\nu}
(\gamma^{\mu\nu} \gamma)_{\alpha\beta} \Big) 
\int\frac{dx_1\,dx_2\, dz d\bar z}{dV_{\rm
CKG}}\,\times \\
&~~~~~~~\times~
\big\langle\,
(S^{\dot \alpha}S^+\ee^{-\varphi/2})(x_1)
(S^{\dot \beta}S^+ \ee^{-\varphi/2})(x_2)
(S^{\alpha}S^- \ee^{-\varphi/2})(z) S^{\beta}S^- \ee^{-\varphi/2}({\bar z}) 
\big\rangle~.
\end{aligned}
\end{equation}
The correlator appearing in the second line above can be obtained by decomposing
the ten-dimensional four-point function of spin fields in $8+2$ dimensions. Due
to the anti-symmetry in $(\alpha\beta)$ of the polarization factor, the only
relevant structure in this correlator is
\begin{equation}
\frac{\frac{1}{2}\,(\gamma^\rho)^{\dot \alpha \alpha}
(\gamma_\rho)^{\dot \beta \beta}}{\big[ (x_1-z)(x_1-{\bar z})
(x_2 -z)(x_2-{\bar z})\big]} ~.
\end{equation}
Then, inserting this into (\ref{corr}) and exploiting the
$\mathrm{Sl}(2,\mathbb{R})$ invariance to fix $x_1\to\infty$ and $z \to \ii$, we
are left with the following elementary integral
\begin{equation}
2 \ii \int_{-\infty}^{\infty}dx_2\,\frac1{1+x_2^2} = 2 \pi \ii~,
\end{equation}
so that, after some algebra, we find
\begin{equation}
\lvev V_{\lambda} V_{\lambda}V_{\cF} \rvev
=
-\frac{1}{16 g_0^2}\, \tr\big\{
\lambda_{\dot\alpha} (\gamma^{\mu\nu})^{\dot \alpha \dot \beta}
\lambda_{\dot\beta}\,\cF_{\mu\nu} \big\}
\label{llF}
\end{equation}
where we have clumped the remaining numerical factors in the normalization of
the background field $\cF$.
With similar calculations, one can compute all other diagrams in Fig.~\ref{fig1}
obtaining
\begin{equation}
\label{DbXF}
\begin{aligned}
\lvev V_{D} V_{\bar \chi}V_{\cF} \rvev
&=
\frac{\ii}{g_0^2}\, \frac{1}{8\sqrt{2}}\,
\tr \big\{D_m (\tau^m)_{\mu\nu} \,{\bar \chi}\, \cF^{\mu\nu} \big\} ~,
\\
\lvev V_{\bar Y} V_{a}V_{\cF} \rvev
&=
\frac{\ii}{g_0^2}\, \frac{1}{2\sqrt{2}}\,
\tr \big\{{\bar Y}_\mu\, a_\nu\, \cF^{\mu\nu} \big\} ~.
\end{aligned}
\end{equation}
Likewise, for the diagrams with the anti-holomorphic background represented in
Fig.~\ref{fig2} we find
\begin{equation}
\label{MMbF}
 \begin{aligned}
 \lvev V_{M} V_{M}V_{\cF} \rvev
 &=
 -\frac{1}{16 g_0^2}\, 
 \tr\big\{ M_{\alpha} (\gamma^{\mu\nu})^{\alpha \beta}M_{\beta}\,
\bcF_{\mu\nu} \big\} ~, \\
 \lvev V_{D} V_{\chi}V_{\bcF} \rvev
 &=
 \frac{\ii}{g_0^2}\, \frac{1}{8\sqrt{2}}\,
 \tr \big\{D_m (\tau^m)_{\mu\nu} \,{\chi}\, \bcF^{\mu\nu} \big\} ~,
 \\
 \lvev V_{Y} V_{a}V_{\bcF} \rvev
 &=
 \frac{\ii}{g_0^2}\, \frac{1}{2\sqrt{2}}\,
 \tr\big\{Y_\mu\, a_\nu\, \bcF^{\mu\nu} \big\}~.
 \end{aligned}
 \end{equation}
 {From} the last lines of Eq.s~(\ref{DbXF}) and (\ref{MMbF}), we see that the
presence of a RR background induces two extra terms in ${\mathcal
S'}_{\mathrm{aux}}$, so that the latter must be replaced according to
 \begin{equation}
  \label{sauxp}
 {\mathcal S'}_{\mathrm{aux}} \to {\mathcal S'}_{\mathrm{aux}} - 
 \frac{\ii}{g_0^2}\, \frac{1}{2\sqrt{2}}\,
 \tr\big\{ {\bar Y}_\mu\, a_\nu\, \cF^{\mu\nu}\,+\, Y_\mu\, a_\nu\, \bcF^{\mu\nu}
  \big\}~.
 \end{equation}
As a consequence, the equations of motion of the auxiliary fields change and
\eq{auxval} must be replaced by 
\begin{equation}
\label{auxval1}
Y_{\mu} =-\left[a_{\mu},\chi\right]\,
 +\, \frac{\ii}{2\sqrt{2}} \,\cF_{\mu\nu} a^\nu~,~~
\bar{Y}^{\mu}=-\left[a^\mu,\bar\chi\right]\,+\, 
\frac{\ii}{2\sqrt{2}}\,\bcF^{\mu\nu} a_\nu~.
\end{equation}
Thus, eliminating $Y_\mu$ and ${\bar Y}^\mu$ we recover the new $\cF$-dependent 
quartic action
\begin{equation} 
\label{sqf}
\begin{aligned}
{\mathcal S'}_{\rm quartic} (\cF, \bcF) =&
{\mathcal S'}_{\rm quartic} +
\frac{1}{g_{0}^2}\, {\rm tr}\, \Big \{ 
\frac{\ii}{2\sqrt{2}}\comm{a_\mu}{\bar\chi}\mathcal F^{\mu\nu}a_\nu  \\
&~+\frac{\ii}{2\sqrt{2}}\comm{a_\mu}{\chi}\bar{\mathcal F}^{\mu\nu}a_\nu 
+ \frac{1}{8}\,\bar{\mathcal F}^{\mu\nu}a_\nu 
{\mathcal F}_{\mu\rho}a^\rho \Big\}~,
 \end{aligned}
\end{equation}
which reproduces the expression given in Eq.s (\ref{squartic2}) and
(\ref{sbarF}) of the main text.

Furthermore, from Eq.s (\ref{llF})-(\ref{MMbF}) we obtain the following
background-dependent cubic terms:
\begin{equation}
\label{scf}
\begin{aligned}
{\mathcal S}_{\rm cubic} (\cF, \bcF) & = 
{\mathcal S}_{\rm cubic} + \frac{1}{g_0^2}\, 
\tr\big\{ \frac{1}{16}\,
\lambda_{\dot\alpha} (\gamma^{\mu\nu})^{\dot \alpha \dot \beta} 
\lambda_{\dot\beta}\,\cF_{\mu\nu} +
\frac{1}{16}\, M_{\alpha} (\gamma^{\mu\nu})^{\alpha \beta}M_{\beta}
\,\bcF_{\mu\nu} \\ 
&~~~~+ \frac{\ii}{8\sqrt{2}} D_m (\tau^m)_{\mu\nu} \,{\bar \chi}\, 
\cF^{\mu\nu} + \frac{\ii}{8\sqrt{2}} D_m (\tau^m)_{\mu\nu} \,{\chi}\,
\bcF^{\mu\nu}\big \}~.
\end{aligned}
\end{equation}
To compare this expression with that used in Section \ref{sec:RRback}, we have
first to decompose the background fluxes as in \eq{deco28} and use the
relabelled fermion moduli defined in \eq{octmod}. Then, performing the traces on
the $\tau$-matrices, one can show that the couplings involving $D_m$ receive
contributions only from the $\cF^{\bf 7}$ and $\bar\cF^{\bf 7}$ components,
given by
\begin{equation}
\label{Dh}
\frac{\ii}{8\sqrt{2}} \cF^{\mu\nu}\, D_m (\tau^m)_{\mu\nu} \,{\bar \chi}=
- \frac{\ii}{\sqrt{2}} h^m D_m \,{\bar \chi}~,~~
\frac{\ii}{8\sqrt{2}} \bcF^{\mu\nu}\, D_m (\tau^m)_{\mu\nu} \,{\chi} =
- \frac{\ii}{\sqrt{2}} {\bar h}^m D_m \,\chi~. 
\end{equation}
On the other hand, using Eq.s~(\ref{tauC}) and (\ref{idtau2gamma2}) we can
rewrite the fermionic bilinears which appear in the first line of \eq{scf} as
follows 
\begin{equation}
\label{llf1}
\begin{aligned}
\frac{1}{16}\,\cF_{\mu\nu}\, \lambda_{\dot\alpha} 
(\gamma^{\mu\nu})^{\dot \alpha \dot \beta} \lambda_{\dot\beta} 
&= \frac{1}{16}\, \cF_{\mu\nu}\, \lambda_m (\gamma^{\mu\nu})^{mn} \lambda_n+
\frac{1}{8} \cF_{\mu\nu}\, \lambda_m (\gamma^{\mu\nu})^{m8} \eta 
\\
& = -\frac{1}{2} \,f^{mn} \lambda_m \lambda_n + h^m \lambda_m \eta
\\
\frac{1}{16}\,\bcF_{\mu\nu}\, M_{\alpha} (\gamma^{\mu\nu})^{\alpha \beta} 
M_{\beta} &= \frac{1}{16}\, \bcF_{\mu\nu}\, 
M^\rho \left(\delta^{\mu\nu}_{\rho\sigma} - C^{+\mu\nu}_{~~~\rho\sigma}\right)
M^\sigma~
\\
& = \frac{1}{8} \,{\bar f}^{mn} (\tau_{mn})_{\rho\sigma} M^\rho M^\sigma - 
\frac{1}{4}\, {\bar h}^m (\tau_m)_{\rho \sigma} M^\rho M^\sigma~~.
\end{aligned}
\end{equation}
{From} these expressions we retrieve the terms of Eq.s (\ref{squartic2}),
(\ref{sbarF}) and (\ref{hterms}) of the main text for $\bar h_m=0$.

\section{Details on the SO$(k)$ integrals}
\label{app:detsok}

\paragraph{Weight sets of SO$(2n+1)$}
This group has rank $n$.
If we denote by $\ve{i}$ the versors in the 
$\mathbb{R}^n$ weight space,
\begin{itemize}
 \item the set of the $2n+1$ weights $\vec\pi$ of the vector representation is 
 given by
 \begin{equation}
  \label{setvec1}
  \pm\ve i~,~~~~ \vec 0~\mbox{with multiplicity $1$}~;
 \end{equation}
 \item the set of $n(2n+1)$ weights $\vec\rho$ of the adjoint representation 
(corresponding to the two-index antisymmetric tensor) is the following:
\begin{equation}
 \label{setrho1}
 \pm \ve i \pm \ve j~(i < j)~,~~~~
 \pm \ve i~,~~~~
 \vec 0~\mbox{with multiplicity $n$}~;
\end{equation}
\item the $(n+1)(2n+1)$ weights of the two-index symmetric tensor%
\footnote{In fact, this is not an irreducible representation: it decomposes 
into the $(n+1)(2n+1)-1$ traceless symmetric tensor plus a singlet. One of the
$\vec 0$ weights corresponds to the singlet.} are
\begin{equation}
 \label{setsymm1}
 \pm \ve i \pm \ve j~(i < j)~,~~~~
 \pm \ve i~,~~~~
 \pm 2 \ve i~,~~~~
 \vec 0~\mbox{with multiplicity $n+1$}~.
\end{equation} 
\end{itemize}

\paragraph{Weight sets of SO$(2n)$}
This group has rank $n$.
If we denote by $\ve{i}$ the versors in the 
$\mathbb{R}^n$ weight space,
\begin{itemize}
 \item the set of the $2n$ weights $\vec\pi$ of the vector representation is 
 given by
 \begin{equation}
  \label{setvec}
  \pm\ve i~;
 \end{equation}
 \item the set of $n(2n-1)$ weights $\vec\rho$ of the adjoint representation 
(corresponding to the two-index antisymmetric tensor) is the following:
\begin{equation}
 \label{setrho}
 \pm \ve i \pm \ve j~(i < j)~,~~~~
 \vec 0~\mbox{with multiplicity $n$}~;
\end{equation}
\item the $n(2n+1)$ weights of the two-index symmetric tensor%
\footnote{Again, this is not an irreducible representation, since it 
contains a singlet.} are
\begin{equation}
 \label{setsymm}
 \pm \ve i \pm \ve j~(i < j)~,~~~~
 \pm 2 \ve i~,~~~~
 \vec 0~\mbox{with multiplicity $n$}~.
\end{equation}
\end{itemize}

\paragraph{SO$(7)$ and its spinorial weights}
The SO$(7)$ rotation group parametrized by the RR fluxes $f_{mn}$ defined in
\eq{Ff} act on the moduli $a^\mu, M^\mu$ 
in its spinorial representation $\mathbf{8}_s$. The set of weights of this
representation is
\begin{equation}
 \label{spinweights}
 \vec\beta = \frac 12 (\pm\ve 1 \pm \ve 2 \pm \ve 3)
\end{equation}
and we define as ``positive'' weights those for which the product of the three
signs is $-1$: 
\begin{equation}
 \label{posspin}
 \begin{aligned}
  \vec\beta_1 & = \frac 12 (-\ve 1 +\ve 2 +\ve 3)~,~~
  \vec\beta_2 = \frac 12 (\ve 1 -\ve 2 +\ve 3)~,~~
  \vec\beta_3 = \frac 12 (\ve 1 +\ve 2 -\ve 3)~,\\
  \vec\beta_4 & = \frac 12 (-\ve 1 -\ve 2 -\ve 3)~,
 \end{aligned}
\end{equation}
so that the combinations
\begin{equation}
 \label{Eascartan}
  E_A = \vec f \cdot \vec\beta_A
\end{equation}
are exactly the combinations introduced in \eq{fE} in the text.

\paragraph{Integration in the cases $k=4,5$}
\label{subapp:k45}
The group SO$(4)$ has rank 2, and the poles of the integrand of \eq{Zkred} are
determined by the polynomial $\cQ(\chi_1,\chi_2)$. According to \eq{Qnulleven}
and to the set of weights in \eq{setsymm}, the $\chi$-dependent part of $\cQ$
(i.e., the one determined from the non-zero weights) is
\begin{equation}
 \label{den4}
 \begin{aligned}
 \prod_{A=1}^4 & (2\chi_1 - E_A)(-2\chi_1 - E_A)
 (2\chi_2 - E_A)(-2\chi_2 - E_A)\\
  & (\chi_1 -\chi_2 - E_A)(-\chi_1 +\chi_2 - E_A)
 (\chi_1 +\chi_2 - E_A)(-\chi_1 -\chi_2 - E_A)~.
 \end{aligned}
\end{equation}
Let's label the various types of monomials from 1 to 8 in the order appearing
above. With the prescriptions given in \eq{imparts}, it is straightforward to
see that all poles in the integrand of \eq{Zkred} are simple (in certain cases,
apparent double poles are compensated by zeroes of the Vandermonde determinant).
We have to sum the residues over different possible classes of poles. For
instance, we could, from the $\chi_1$ integral, pick up the residue from a
simple pole determined by the 5${}^{\mathrm{th}}$ factor in \eq{den4}:
\begin{equation}
 \label{example1}
 \chi_1 = \chi_2 + E_A~.
\end{equation}
After substituting this value in the remaining terms of the integrand, we 
integrate over $\chi_2$ and we can again pick up contributions from various
possible poles. For instance, suppose that we choose the one coming from the
third factor:
\begin{equation}
 \label{example2}
 \chi_2 =  \frac{E_B}{2}
\end{equation}
and make this replacement in all remaining factors of the integrand to compute
the residue. The choices \eq{example1} and \eq{example2} are possible for all
$A,B$, so we have to sum the residues over $A,B$ independently; let us write
this particular contribution to the integral as
\begin{equation}
 \label{excont}
 \sum_{A,B} (5,3)~.
\end{equation}
With this condensed notation, it is straightforward to check that the
contributions to the integral are the following: 
\begin{equation}
 \label{contk4}
 \sum_{A\not= B} (1,3) + \sum_{A,B} (1,6) + \sum_{A,B} (5,3) + \sum_{A,B} (7,2) +
 \sum_{A\not=B} (7,3) + \sum_{A,B} (7,6)~.
\end{equation}
In fact, there are also other contributions that, however, cancel in pairs:
\begin{equation}
 \label{contcanc4}
 0 = \Bigl(\sum_{A\not= B} (1,7) + \sum_{A\not= B}  (7,1) \Bigr) +
\Bigl(\sum_{A > B} (5,7) + \sum_{A < B}  (7,5) \Bigr) ~. 
\end{equation}
Evaluating explicitly the sums in \eq{contk4} one obtains $Z_4$; inserting it in
\eq{Fkexp} one determines $F_4$, as described in the main text. 

Let us now move to SO$(5)$, which again has rank 2. According to \eq{Qnullodd}
and to the set of weights in \eq{setsymm1}, the $\chi$-dependent part of $\cQ$
(i.e., the one determined from the non-zero weights) is 
\begin{equation}
 \label{den5}
 \begin{aligned}
 \prod_{A=1}^4 & (2\chi_1 - E_A)(-2\chi_1 - E_A)
 (2\chi_2 - E_A)(-2\chi_2 - E_A)\\
 & (\chi_1 -\chi_2 - E_A)(-\chi_1 +
 \chi_2 - E_A)(\chi_1 +\chi_2 - E_A)(-\chi_1 -\chi_2 - E_A)\\
 & (\chi_1 - E_A)(-\chi_1 - E_A)(\chi_2 - E_A)(-\chi_2 - E_A)~.
 \end{aligned}
\end{equation}
Let us label the various types of monomials from 1 to 12 in the order appearing
above. One can check that only simple poles appear and, using the condensed
notation introduced above, the classes of residues that contribute to $Z_5$ are
the following:
\begin{equation}
 \label{contk5}
 \begin{aligned}
  & \sum_{A\not= B} (1,3) + \sum_{A,B} (1,6) +  \sum_{A,B} (1,11) +
    \sum_{A,B} (5,3) + \sum_{A,B} (5,11) + \sum_{A,B} (7,2) + \sum_{A\not=B} (7,3) \\ 
+ & \sum_{A,B} (7,6) + \sum_{A,B} (7,10) + \sum_{A\not= B} (7,11)  + \sum_{A,B} (9,3) + 
    \sum_{A,B} (9,6) + \sum_{A\not= B} (9,11)~,
  \end{aligned}
\end{equation}
having already taken into account the pairwise cancellation of other classes of
contributions:
\begin{equation}
 \label{contcanc5}
 \begin{aligned}
 0 & = \Bigl(\sum_{A\not= B} (1,7) + \sum_{A\not= B}  (7,1) \Bigr) +
       \Bigl(\sum_{A > B} (5,7) + \sum_{A < B}  (7,5) \Bigr) \\
   & + \Bigl(\sum_{A > B} (5,9) + \sum_{A < B}  (9,5) \Bigr) +
       \Bigl(\sum_{A < B} (7,9) + \sum_{A > B}  (9,7) \Bigr)~. 
 \end{aligned}
\end{equation}
Explicitly evaluating the sums in \eq{contk5} one obtains $Z_5$; inserting it
in \eq{Fkexp} one determines $F_5$, as described in the main text.

\providecommand{\href}[2]{#2}\begingroup\raggedright

\endgroup

\end{document}